\documentclass[ ]{aa}

\usepackage{hyperref} 
\usepackage{graphicx}
\usepackage{color}
\usepackage{nicematrix}
\usepackage[varg]{txfonts}
\usepackage{multirow}
\usepackage{makecell}
\usepackage{amsmath}
\usepackage{rotating}
\usepackage{gensymb}
\usepackage{threeparttable}
\usepackage[round]{natbib}
\title{Evidence of global periodicity in PSP data:\\ can they be linked to long-period oscillations in solar active regions?}

\author{G.~Dumbadze\inst{1,2,3,5}, B.M.~Shergelashvili\inst{1,2,3,4}, M.~Khodachenko\inst{4}, L.~Westrich\inst{1,2}, H.~Fichtner\inst{1}, A.~Reza\inst{4} and S.~Poedts\inst{5,6} }

\institute{Institute for Theoretical Physics IV, Ruhr-Universität Bochum, Universitätsstrasse 150, 44780 Bochum, Germany\\
                \and
         Centre for Computational Helio Studies, Faculty of Natural Sciences and Medicine, Ilia State University, Cholokashvili Ave. 3/5, 0162 Tbilisi, Georgia\\
                  \and
         Evgeni Kharadze Georgian National Astrophysical Observatory, M. Kostava street 47/57, 0179 Tbilisi, Georgia\\
                 \and
         Institut f\"ur Weltraumforschung, \"Osterreichische Akademie der Wissenschaften, Schmiedlstrasse 6, 8042 Graz, Austria\\
                \and
        Centre for Mathematical Plasma Astrophysics/Department of Mathematics, Celestijnenlaan 200B, 3001 Leuven, Belgium\\
                 \and
        Institute of Physics, University of Maria Curie-Skłodowska, Pl.\ Marii Curie-Sk{\l}odowskiej 1, 20-031 Lublin, Poland
}

\begin{document}

%%%%%%%%%%%%%%%%%%%%%%%%%%%%%%%%%%%%%%%%%%%%%%%%%%%%%%%%%%%
\abstract
{The presence of coherent oscillatory signatures in Parker Solar Probe SWEAP datasets, identified through a series of spectral analyses, is addressed, and their possible connection to oscillatory phenomena in HMI/SDO datasets of active regions is questioned. }
{The goal of the work is to perform a systematic analysis of oscillation spectra applied to SWEAP/PSP time series of various physical quantities. }
{We apply the Lomb-Scargle spectral analysis method to the original, unevenly sampled PSP datasets and to wavelet analyses corresponding to the interpolated, evenly sampled time series. Furthermore, the statistical analysis uses spectra developed for 17 PSP encounters (1-2, 4-9, 11-19) to obtain statistically proven characteristic periodicities.}
%-
%{It has been found that, similarly to previously studied spectra in solar active regions with periods above 2 hours, the solar wind datasets manifest sequences of statistically significant periods in the same range in datasets of different physical quantities. }
%-
{It has been found that, similarly to previously studied spectra in solar active regions with periods exceeding 2 hours, the solar wind datasets exhibit sequences of statistically significant periods in the same range across datasets of different physical quantities.}
%-
{Based on the analysis, it is concluded that the spectra exhibit coherent, fixed periodicities in the range of 2-20 hours, with the highest statistical confidence in periods between 4 and 8 hours. It is also vital that the signature of such periods is systematically absent in datasets of PSP orbital trajectory far from the encounter sites. The coherent signal is detectable at the place of the orbital perihelion (when PSP moves almost horizontally with respect to the solar surface) and during approaching and receding phases when PSP flies almost radially. Discrepancies in measured periods across different physical quantities may indicate distinct oscillatory driving mechanisms.}

%%%%%%%%%%%%%%%%%%%%%%%%%%%%%%%%%%%%%%%%%%%%%%%%%%%%%%%%%%%

\titlerunning{Evidence of global periodicity in PSP data}

\authorrunning{Dumbadze, Shergelashvili et al.}

\maketitle
%%%%%%%%%%%%%%%%%%%%%%%%%%%%%%%%%%%%%%%%%%%%%
\section{Introduction}\label{intro}
%%%%%%%%%%%%%%%%%%%%%%%%%%%%%%%%%%%%%%%%%%%%%
Understanding the causal relationships between the Sun and the inner/outer heliosphere within a unified framework for physical processes is a cornerstone of space science, as it concerns the Sun's influence on its planetary system. This influence is represented through quasi-stationary and transient patterns of space weather. The key aspect of the scientific interest encompasses the source inner boundary of the solar wind. The Parker Solar Probe (PSP, \citet{Fox2016}) mission aims to observe the part of the heliosphere that plays a role in the origin and acceleration of the solar wind to supersonic and super-Alfvénic velocities, as observed at 1~AU. The periodic approaches of PSP spacecraft to the Sun with unprecedentedly close (down to about 10--11 solar radii) approaches (usually called encounters) enable the measurement of the physical conditions in the vicinity of the solar atmosphere and the solar wind source region. A similar task is fulfilled by the Solar Orbiter mission \citep{SolarOrbiter2020}. There are some significant results related to fluctuation spectra of plasma physical quantities in the solar wind, obtained by the PSP \citep[e.g. see,][and references therein]{Nemecek2021,Wu2021,Krupar2020} mission. 

The content of this work is linked to the characteristic periodicities reported in analyses of datasets from the lower solar atmosphere of other observational missions, which we cite in chronological order. In this respect, it is essential to note that Solar Dynamic Observatory (SDO) datasets clearly show signatures of oscillations both at sub-hour periods \citep{Zhong2025} and within the more extended multi-hour range \citep{Wisniewska2024,Dumbadze2021,Dumbadze2017}.  In all these works datasets from HMI/SDO photospheric magnetograms have been analysed. A thorough analysis of white-light images obtained by the WISPR/PSP instrument within the range of heliocentric distances 8-16$R_{\sun}$, revealed the presence of clearly detected periodicities in the range 1.5--3 hours, and periodicities $>4$ hours of timescale (with characteristic spatial scales as small as 0.1$R_{\sun}$) have been reproduced through the simulations the latter being in good agreement with WISPR and other observations \citep[see,][and references therein]{Poirier2023}. The other works link the detected observational evidence to the presence of relatively long 10--20 hours and short 1--2 hours coherence in the slow solar wind, attributing it to the influence of periodically released flux ropes on the heliospheric current sheet \citep{Reville2020,Reville2022}. We perform a similar study using in situ Solar Wind Electrons Alphas and Protons (SWEAP/PSP, \citet{Kasper2016}) instrument measurements. In the context of density profile modification before and during Coronal Mass Ejection (CME) propagation, this has also been studied in recent years through analyses of type-III radio burst statistics \citep[][and related works]{Dididze2019} and other dynamic processes \citep{Melnik2015}. In general terms, these results can be formulated in conceptual narratives that state that, near its origin, the solar wind exhibits a much more complex dynamics than previously known. Another fundamental finding is a correlation between processes in the solar interior and in the outer atmosphere. The occurrence of dynamical events, in general, follows stochastic behaviour characterised by near- or far-from-equilibrium statistical and thermodynamic regimes. Good examples of such correlated processes are solar flares and magnetic reconnection-based other eruptive events (e.g., destabilised feet of prominences due to the solar tornadoes shown in \citet{Mghebrishvili2018}) can ignite large-scale transient distortions of the solar wind patterns (CMEs). 

Apart from this eruptive ``chaos", another aspect of the issue concerns the quasi-stationary part of the solar wind, which exhibits spatial and temporal coherence. This coherence is manifested through various structures that exhibit a certain spatiotemporal invariance. Moreover, the solar atmosphere and wind exhibit numerous coherent periodicities and oscillatory patterns. These coherence features maintain mutual spatial and/or temporal modulation among quasi-periodic oscillatory processes in the solar interior, the atmosphere, and the wind outflow. Based on extensive observational evidence and modelling, the intuitive assumption is that the core agents mediating correlations between physical processes across various levels are the magnetic field and flow structural patterns, and in certain cases MHD turbulence within them \citep[e.g. see,][]{Bruno2013}.

Recently, the existence of characteristic periodicity in the shapes and magnetic-flux fluctuations of solar active regions has been reported \citep{Dumbadze2021,Dumbadze2017}. The spectrum of typical periods spans from 2 to 20 hours. However, the main power — the most significant statistical frequency of occurrence — has been detected at about 4.5 and 6-hour periods. Specific spectral analysis techniques have been applied to distinguish these periodicities from intense background noise. It should be noted that the significant power of these oscillations was measured only in active-region magnetic-field-concentration areas. In contrast, in quiet Sun regions populated by spurious small-scale magnetic fields sitting between the supergranulation cells, the signatures of long-periodic coherent oscillations vanish. The main conclusion of these studies is that the observed oscillations are attributed solely to active-region structures. This observational fact provokes the question: \textit{are these oscillations the local attributes of active region structural vibrations, or they also influence the upper large-scale areas in the solar corona and inner heliosphere?}

Moreover, typical coherent periodicities have recently been detected in PSP datasets as well \citep{Bale2021}. They are observed during encounters near the Sun while the PSP spacecraft passes along its orbital trajectory at perihelion. This means that it is part of the orbit locally perpendicular to the radial direction. Based on this circumstance, it has been concluded by the authors that the coherent periodicities are most plausibly manifestations of passing of PSP through the longitudinally and latitudinally extended coherent spatial structures in the near vicinity of the orbital perihelion linked with supergranulation cells. This assumption inspired the other inquiry: \textit{could this conclusion be properly confirmed by comparing periodograms obtained from datasets recorded when PSP flies almost radially in the heliocentric frame, before (inbound flight) and after (outbound flight) the perihelion flyby?}

The work presented in this manuscript addresses two issues mentioned above in combination to examine the eligibility of the hypothesis that, {\it the oscillations seen in the active regions and those found by PSP mission in the young solar wind datasets might be mutually linked, representing two manifestations of a standard oscillatory dynamic process.} 

To achieve this goal, we perform an extended analysis of the SWEAP/PSP in situ measurements, focusing not only on the vicinity of the PSP orbital perihelion but also on the flight epochs of the encounters, when the spacecraft approaches and recedes from the Sun nearly radially. It should be noted that we analyse in this work the Solar Probe Cup (SPC) part of the SWEAP/PSP datasets. Under the almost radial approaching (receding), we mean the $\pm 4^{\degree}$ heliocentric longitudinal intervals of the encounter trajectories around the points where the longitudinal angular velocity of PSP $\partial \varphi /\partial t$ vanishes. At the same time, to reduce contaminating edge effects, the analysed data sets for each epoch are obtained with additional external margins of approximately $10\%$ of the considered timespan, defined separately in each case (see below). In addition, we also analyse data obtained between different encounters, far from the Sun at distances $>100~R_{\sun}$. In total, we analysed the datasets for 17 PSP orbital encounters (from encounter 1 to 19, excluding encounters 3 and 10, for which there were insufficient data samples for our analysis). 

As the PSP changes the orbital parameters between encounters, the spacecraft's actual time spans and radial velocities are adjusted accordingly. Indeed, the local plasma environment conditions along different encounter trajectories can differ; nonetheless, the applied subdivision of the encounter into quasi-radial and horizontal (perihelion) motion with respect to the solar surface allows reliable probing of the large-scale spatiotemporal modulations in the corresponding regions of the inner heliosphere.  
Our ultimate goal is to address the following issues: 
\begin{itemize}
    \item Are the long-period oscillations reported for active region SDO/HMI magnetograms \citep{Dumbadze2021,Dumbadze2017} the attributes solely of the subsurface and solar atmospheric magnetic structure of the active regions or the signatures of similar coherent signals can be found more globally in the upper solar atmosphere and surrounding wind stream?
    \item Could a similar set of significant periodicities be found aside from the perihelion along the almost radial parts of the orbital trajectory on both approaching and receding sides? The first indication of such signatures during the PSP horizontal flight across atmospheric magnetic structures near the orbit perihelion was reported by \citet{Bale2021}. 
    \item Are these signatures seen in time series for different physical quantities obtained through the in-situ measurements of proton velocity distributions and related integral moments, made by the SWEAP/PSP instrument? 
    \item Is there a drastic discrepancy in signatures of the coherent fluctuations between close-by encounters and far away from the Sun parts of the PSP orbit (distances $>100~R_{\sun}$)? 
\end{itemize}

The manuscript is organised as follows. In the second section, we present the methodological approach for elaborating on the original datasets and generating periodogram spectra. As the original timeseries are unevenly spaced, we use the Lomb-Scargle spectral technique rather than the standard Fast Fourier transform. We also apply the Continuous Wavelet Transform (CWT) to the datasets; however, in this case, the original, unevenly spaced time series are first interpolated to a constant sampling rate. In the third section, we present the statistical analysis results, with a corresponding discussion in the fourth section, for the significant periods identified in the periodograms using a well-defined criterion. In the last section, some essential conclusive remarks are given.  
%%%%%%%%%%%%%%%%%%%%%%%%%%%%%%%%%%%%%%%%%%%%%%%%
\section{Methodological approach}\label{method}
%%%%%%%%%%%%%%%%%%%%%%%%%%%%%%%%%%%%%%%%%%%%%%%%
We consider the datasets of the SWEAP/PSP instrument when it was at its closest, at the orbital perihelion (encounter) and the farthest (i.e., for distances $>100~R_{\sun}$, including aphelion) position to the Sun. We applied the procedures of data spectral (Lomb-Scargle and CWT) analysis to the proton number density $n_p$, the proton bulk radial velocity $V$ and temperature $T=m_p V_\mathrm{th}^2/2k_\mathrm{B}$ (calculated from observed thermal velocity), which are directly provided by SWEAP/PSP, as well as to the derived quantities, such as normalised mass flux rate $M_f=n_p V R_\mathrm{PSP}^2$; Specific entropy $S \sim T/n_p^{2/3}$ assuming that adiabatic index is $5/3$; and pressure $p=n_p k_\mathrm{B} T$. Above, we used the notations: $m_p$ is the proton mass and $k_\mathrm{B}$ is the Boltzmann constant. The available in SWEAP/PSP data archive values of $n_p$, $V$ and $V_\mathrm{th}$ (equivalently the temperature $T$), calculated via moments of the proton velocity distribution function (VDF), are obtained there in two different ways, depending on the method for the reconstruction of the distribution function. In the first case, which we refer to as the Direct Moment Method (DMM), the VDF is reconstructed from directly measured proton velocities. In the second approach, called the Fitted Moment Method (FMM), the observed proton VDFs are fitted with a single or multiple Maxwellian functions. It is worth noting that, in many cases, the observed distributions are substantially non-Maxwellian, indicating that the plasma deviates from thermodynamic equilibrium. 

The availability of data of interest was the primary criterion for selecting particular PSP orbital encounters for analysis, specifically Nrs.\ 1,2, 4--9, 11--19. However, even during these encounters, the particle VDFs (directly measured and fitted) and the corresponding reconstructed physical quantities were not always available in the SWEAP/PSP instrument's database. Apart from these generic gaps in datasets there are a significant number of SPC dataset quality flags indicating low rate of reliability of corresponding parts of data. For the sake of maximum efficiency of our spectral analysis, we discarded all the data standing under these quality flagships. This particular concern, the perihelion phases of the encounters, whereas the data sets for the approaching and receding phases are more complete. The reason for this difference is that, during perihelion phases, the onboard instruments were often switched off due to substantial solar impact (especially after encounter 9) or datasets were vastly not reliably recorded after 6th encounter, as shown in Tables~\ref{tableD}-\ref{tableT}. In contrast, during the approaching and receding phases, data were recorded more continuously with relatively few gaps. It is also worth noting that datasets for different physical quantities exhibit additional, even non-overlapping, gaps. This is related to the specifics of the onboard calculation of the different-order moments of the reconstructed VDFs. In particular, the zeroth order moments, which provide the density, are contributed mainly by the central parts of VDFs, while the higher order (1st and 2nd) moments are more dependent on the tails of the VDFs. It can even happen that the integrals for the different order moments based on the reconstructed VDFs are not always simultaneously convergent. Consequently, the datasets at various moments and their combinations do not coincide. To detect periodic modulations in the considered physical quantities from these partially incomplete data, while mitigating the adverse effects of gaps, we base our approach on averaging the obtained fluctuation spectra over all considered encounters and on VDF reconstruction methods, and consequently evaluate the statistical significance of the detected periodicities.

\begin{figure*}[ht!]%
\centering{\includegraphics[width=1.0\textwidth]{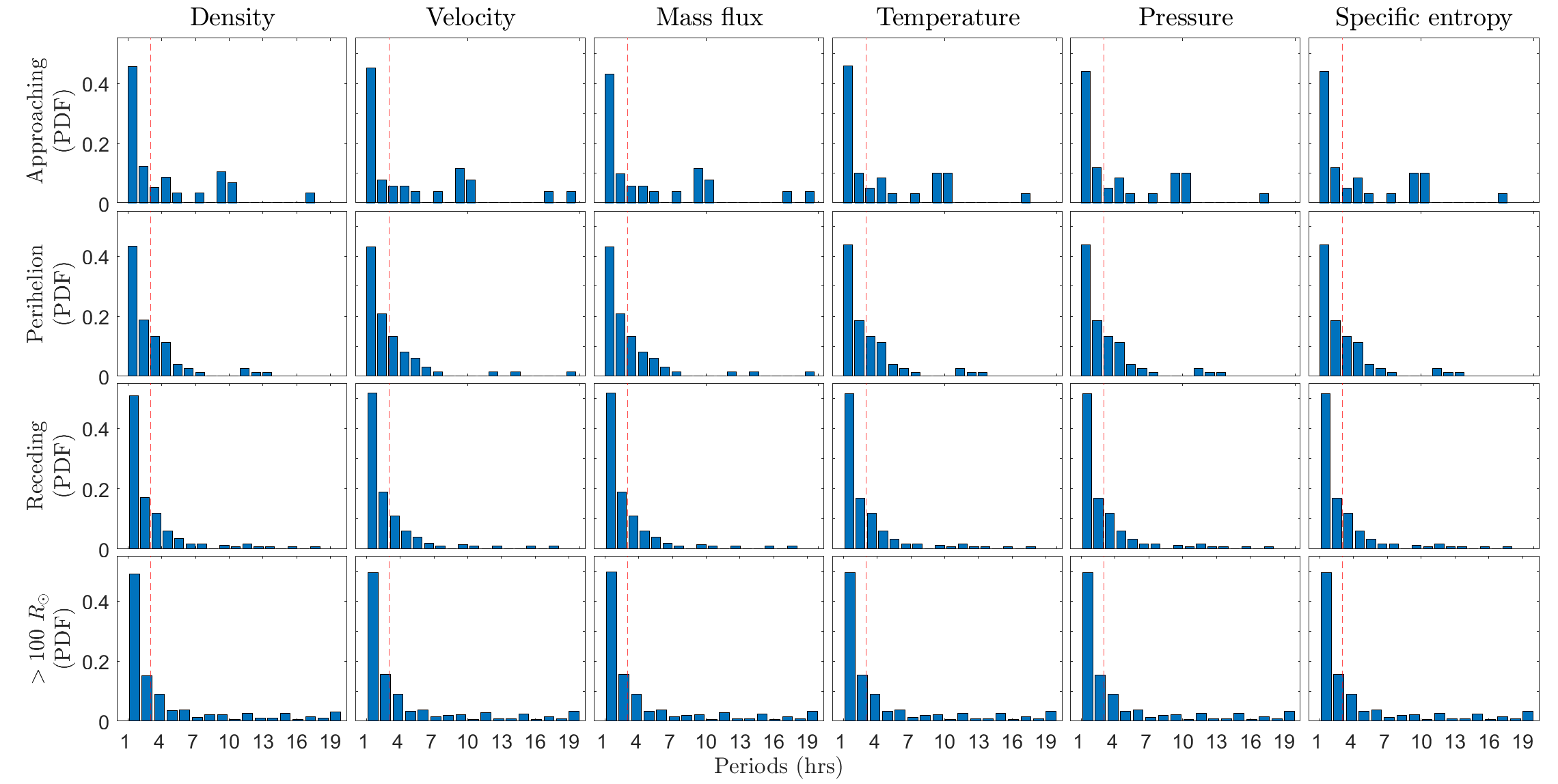}}
\caption{Normalized occurrence frequency
histograms (representing Probability Distribution Functions (PDF)) with 1-hour bin size of the dataset gap relative occurrence frequencies over the set of their durations. The histograms (the example sets of discovered total sum duration of all gaps in the density, bulk radial velocity and temperature datasets are given in the Appendix Tables~\ref{tableD},~\ref{tableV} and \ref{tableT}, respectively) are plotted for all considered physical quantities and flight epochs. The part of the histogram up to 3 hours is indicated by the vertical red dashed lines in all panels.
\label{gaps}}
\end{figure*}

The detailed overview of the availability of analysed SWEAP/PSP data over considered encounters and flight epochs, as well as applied VDF reconstruction methods, with the indication of the relative portion of the gaps, is provided (see, Appendix) in Tables \ref{tableD} -- \ref{tableT}, for  $n_p$ (VDF's zeroth-order moment), $V$ (VDF's first-order moment) and $T$, expressed via $V_\mathrm{th}$ (VDF's second-order moment), respectively, and in Fig.~\ref{gaps} we present histograms of the gap occurrence rates over the gap duration, sampled at 1-hour bin size. The maximum occurrence probability is for gaps with durations in the interval 1--2 hours, and next in this statistical list are gaps lasting 2--3 hours (the part of the histogram up to 3 hours is indicated by the vertical red dashed lines in all panels of Fig.~\ref{gaps}). For some physical quantities, several relatively statistically insignificant yet noticeable gaps of longer duration are also observed. The presence of data gaps at all these scales, and especially in the first two ranges, may affect the obtained spectra (depending on the applied method) and therefore must be adequately accounted for. In the course of analysis, we treated these data gaps differently to produce the Lomb-Scargle and CWT spectra. More details on this are provided below in the subsections \ref{met_LS} and \ref{met_wavelet}, respectively. It is also worth noting that, as will be shown later, despite different treatments of the gaps, the statistically significant results of the performed Lomb-Scargle and CWT analyses were essentially the same, indicating the sufficient quality of the data and consistency of the applied analysis methods.

%%%%%------figure of histogram of gaps
%%%%%%%%%%%%%%%%%%%%%%%%%%%%%%%%%%%%%%%%%%%%%%%%
%%%%%%%%%%%%%%%%%%%%%%%%%%%%%%%%%%%%%%%%%%%%%%%%
Given the partial field of view of the SWEAP/PSP instruments, which may cause discrepancies between the reconstructed and real distribution functions and result in specific mutual dependencies of their momenta, we aim to study multiple encounters and their phases to gain sufficient statistics for the detected quasi-periodic phenomena of interest and to avoid nonphysical artifacts. In our analysis, we use the Lomb-Scargle periodogram and the Continuous Wavelet Transform (CWT) to investigate the quasi-oscillatory spectra in the fluctuations of the measured physical quantities. Both methods are applied to all the flight epochs specified above, i.e., the approaching phase (inbound), perihelion phase, receding phase (outbound) and far flights (beyond $100~R_{\sun}$).

%%%%%%%%%%%%%%%%%%%%%%%%%%%%%%%%%%%%%%%%%%%%%%%%
\subsection{Method 1: Lomb-Scargle power spectra}\label{met_LS}
%%%%%%%%%%%%%%%%%%%%%%%%%%%%%%%%%%%%%%%%%%%%%%%%
The Lomb-Scargle periodogram method  \citep{VanderPlas2018} allows the detection and testing of periodic signals of interest in unevenly spaced data. We chose this spectral analysis method over the standard fast Fourier transform because the SWEAP/PSP data are sampled with irregular time steps. Moreover, we perform the Lomb-Scargle analysis on the merged datasets, with all gaps removed and all consecutive data portions matched. This shrinks the total duration of the analysed time spans accordingly. Such an approach is typically applied to the spectral analysis of fragmented time series from laboratory and space data before applying Fourier-type transforms. Therefore, without loss of generality and any damage to the detected spectra, we do not keep information about the original data timeline because the Fourier-type transform we use produces only the frequency (periodogram) spectra. 

\begin{figure*}[ht!]%
\centering{\includegraphics[width=1.0\textwidth]{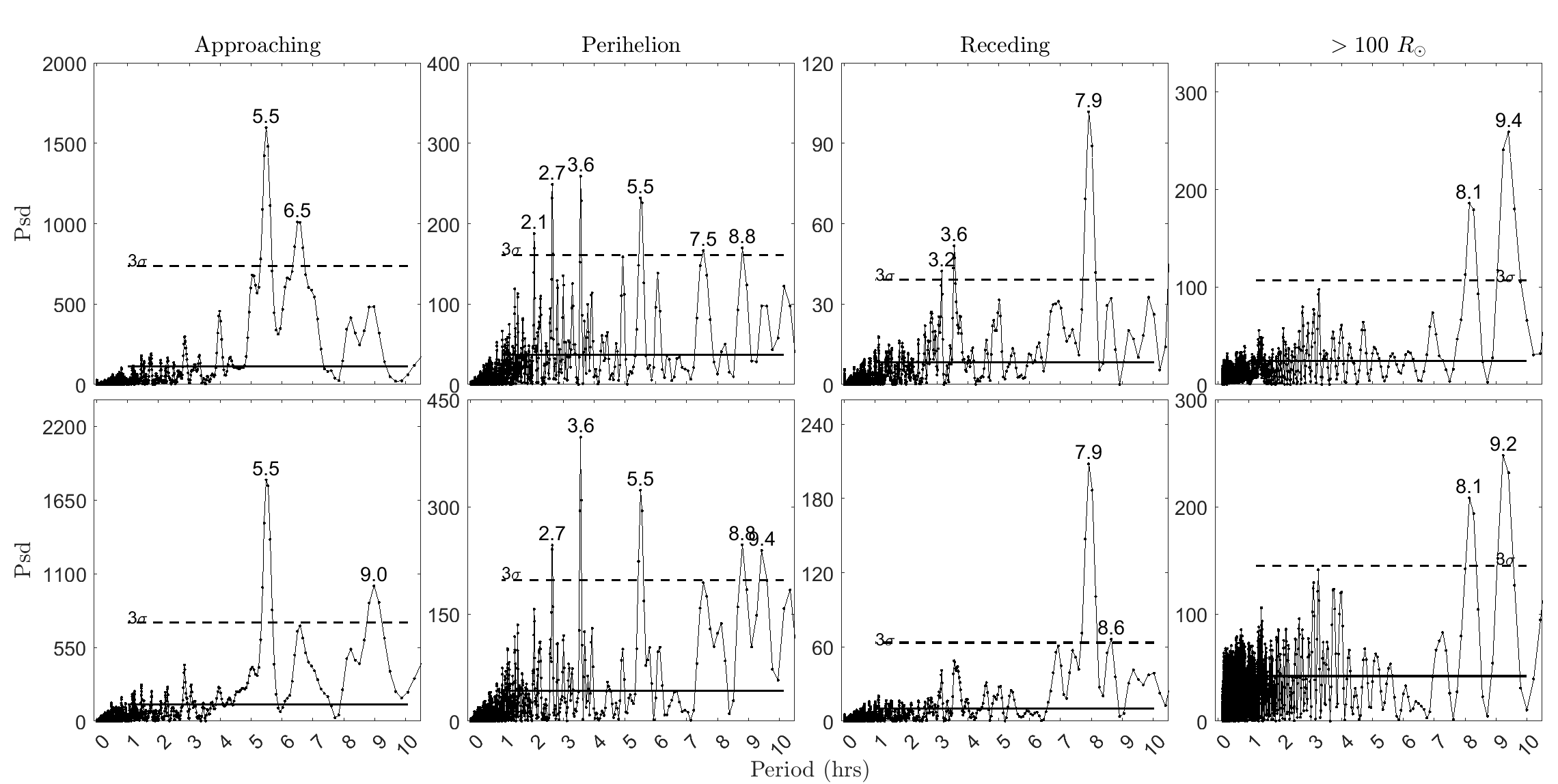}}
\caption{Example of a Lomb-Scargle periodogram for proton number density detected with DMM (upper row of panels) and FMM (bottom row of panels) for the encounter 1. The panels, arranged from left to right, correspond to three parts of the encounter (approaching, perihelion, receding phases) and to the far flight ($>100~R_{\sun}$) trajectories. The horizontal dashed lines indicate levels of $3\sigma$ confidence thresholds (99\%) for the periods in the range 1--10 hours, over the mean of the spectrum depicted with solid horizontal lines. All the spectral power within $3\sigma$ threshold is assumed to be the noise. \label{LS0}}
\end{figure*}

The results of the analysis are presented in Fig.~\ref{LS0}. We compute Lomb-Scargle periodograms for each considered epoch of the encounter separately, as well as for the corresponding far flights. 11-day timespan of the given encounter that we consider encompasses the approaching almost radial, perihelion flyby and receding almost radial phases, altogether. Therefore, we naturally assume that the perihelion flyby phase divides the entire timespan into two approximately equal parts. Correspondingly, for the analysis of the perihelion epoch, we use data measured between the times when the spacecraft's longitudinal angular velocity vanishes ($\partial \varphi /\partial t \sim 0$). For the far flights, the data measured at distances $>100~R_{\sun}$ are taken as a whole (including aphelion) without any additional subdivision. When identifying particular periods in the spectrum within the range of 1–10 hours, we assume with a sufficient degree of confidence that the random part of the spectrum can be approximated within the considered range of periods as white noise. Based on that, in all the panels of Fig.~\ref{LS0}, we indicate with the horizontal dashed lines the corresponding levels of $3\sigma$ confidence thresholds (99\%) for the periods in the range 1--10 hours, where $\sigma = (\sum_n (X_n - <X>)^2 / N)^{1/2}$, N is the sample size of the spectral datasets and $<X> = (\sum_n X_n) / N$ is the sample mean (mean power), denoted with the horizontal solid lines. As can be seen in the panels of Fig.~\ref{LS0}, several peaks appear to exceed the threshold level, indicating they are true outliers and form a coherent part of the spectrum, possibly related to specific quasi-periodic processes in the plasma background of PSP. The final proof of their significance will be based on the statistical analysis of the available measurement data.    

It is worth noting that, in most of the analysed spectrum cases corresponding to inbound, perihelion, and outbound flights at the encounters, several similar spectrally significant peaks with power exceeding $3\sigma$ are observed, indicating that they substantially exceed the background noise level. It should be noted that these peaks appear in the spectra independently of the reconstruction method, i.e., for DMM and FMM (upper and lower rows of panels in Fig.~\ref{LS0}, respectively), with relatively high signal-to-noise ratios. In contrast, the spectra for the far flight epochs (beyond 100 solar radii from the Sun), show a noisier character, despite some peaks above $3\sigma$. However, as shown in the following, these peaks in the far-flight spectra are spurious and statistically insignificant, with signal-to-noise ratios that appear to be systematically low. Moreover, the long-periodic ($8.6$-hour and $10$-hour) peaks that appeared in the far flight density-fluctuation spectrum revealed by the DMM are absent in the analogous spectrum based on the FMM, indicating their unnatural character.  

%%%%%%%%%%%%%%%%%%%%%%%%%%%%%%%%%%%%%%%%%%%%%%%%
%%%%%%%%%%%%%%%%%%%%%%%%%%%%%%%%%%%%%%%%%%%%%%%%
\subsection{Method 2: Continuous wavelet spectra}\label{met_wavelet}
%%%%%%%%%%%%%%%%%%%%%%%%%%%%%%%%%%%%%%%%%%%%%%%%
Another method of data analysis we use to process unevenly sampled time series is the standard MATLAB package for Continuous  Wavelet Transform (CWT). We use it in combination with the Lomb-Scargle spectral method to cross-validate the obtained results. As a preparatory step for performing CWT analysis, we resampled the unevenly spaced datasets of the analysed physical quantities to a constant sampling rate by applying linear interpolation to the original time series. Additionally, we must preserve the original dataset timelines without removing existing gaps. We fill these gaps using the same linear interpolation we applied to produce evenly spaced datasets from the original irregularly spaced time series. Consequently, we obtained a set of evenly sampled datasets, ready for subsequent wavelet analysis.

For each considered encounter data set, we performed the following actions: 
\begin{enumerate}
\item We again distinguish in each encounter its three significant parts: approaching, perihelion, and receding, as specified above, and perform CWT analysis of the available records of physical quantities using two types of wavelet functions: Morse and Bump wavelets. A similar CWT analysis is also performed on the data measured during the far flights ($100~R_{\sun}$).
\item To construct wavelet spectra for the approaching, perihelion, and receding parts of the trajectory, we proceed with two different approaches:
\begin{itemize}
    \item Procedure (i) We construct the CWT spectrum for the entire 11-day time span of the encounter. Furthermore, we cut the resulting spectrum into three consecutive parts, corresponding to the approaching, perihelion, and receding epochs.  
    \item Procedure (ii) We construct the CWT spectra for each considered epoch separately by similarly dividing the analysed data as for the Lomb-Scargle periodogram method.  
\end{itemize}
\item To get an overall view of all detected quasi-periodic signatures, for each CWT spectrogram and considered epoch, we calculate time averages of the corresponding CWT spectra over time and plot them separately as average (global) wavelet spectra. 
\end{enumerate}

\begin{figure*}[ht!]%
\includegraphics[width=1.0\textwidth]{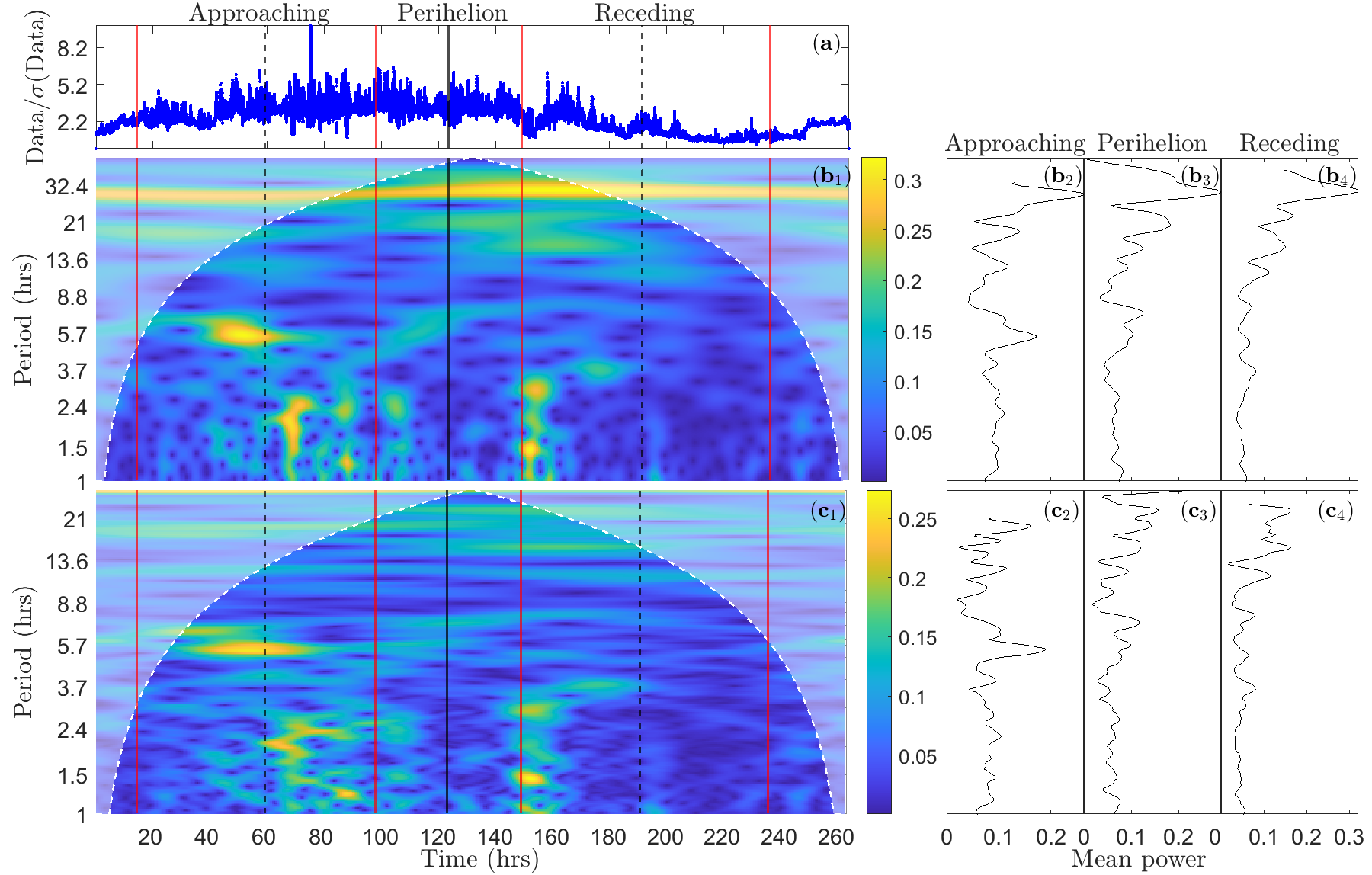}
\caption{Example of CWT analysis of the interpolated normalised dataset for the fluctuations of density obtained with DMM during Encounter 1, within the Procedure (i) (the entire 11-day timespan). In panel (a), the processed dataset is shown. The row of panels (b$_1$,b$_2$,b$_3$,b$_4$) corresponds to Morse and the bottom row (c$_1$,c$_2$,c$_3$,c$_4$) to Bump mother wavelet, respectively. The colour charts (b$_1$,c$_1$) show the dynamical CWT spectra, and the plots to the right (b$_2$ - b$_4$ and c$_2$ - c$_4$) depict their time averages. The faded area in the wavelet spectrum indicates low confidence. The solid black vertical line in the middle indicates the moment of the PSP's perihelion flyby, whereas dashed black vertical lines on both sides of it denote the moments when the spacecraft's orbital motion is 'purely' radial, i.e., its longitudinal angular velocity vanishes ($\partial \varphi /\partial t \sim 0$). Correspondingly, the red vertical lines on the left and right halves of the diagram denote the $\pm 4^{\degree}$ heliocentric longitudinal intervals of the almost radial approaching and receding epochs. The interval between the inner red vertical lines is the perihelion phase.
\label{Enc1-DMM-1}}
\end{figure*}

\begin{figure*}[ht!]%
\includegraphics[width=1.0\textwidth]{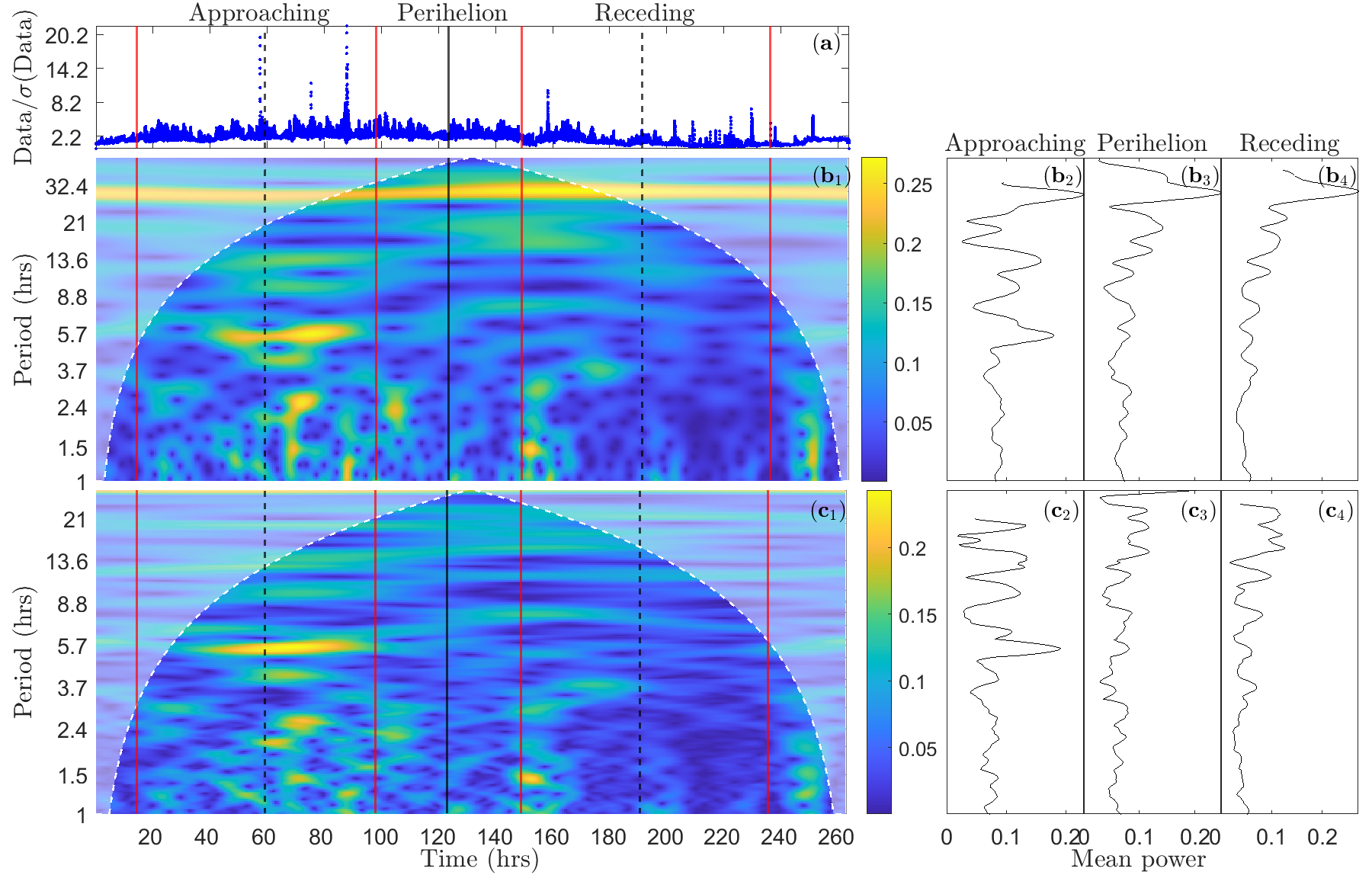}
\caption{Same as in Fig.~\ref{Enc1-DMM-1}, for the fluctuations of density obtained with FMM. \label{Enc1-FMM-1}}
\end{figure*}

\begin{figure}[ht!]%
\includegraphics[scale=0.33]{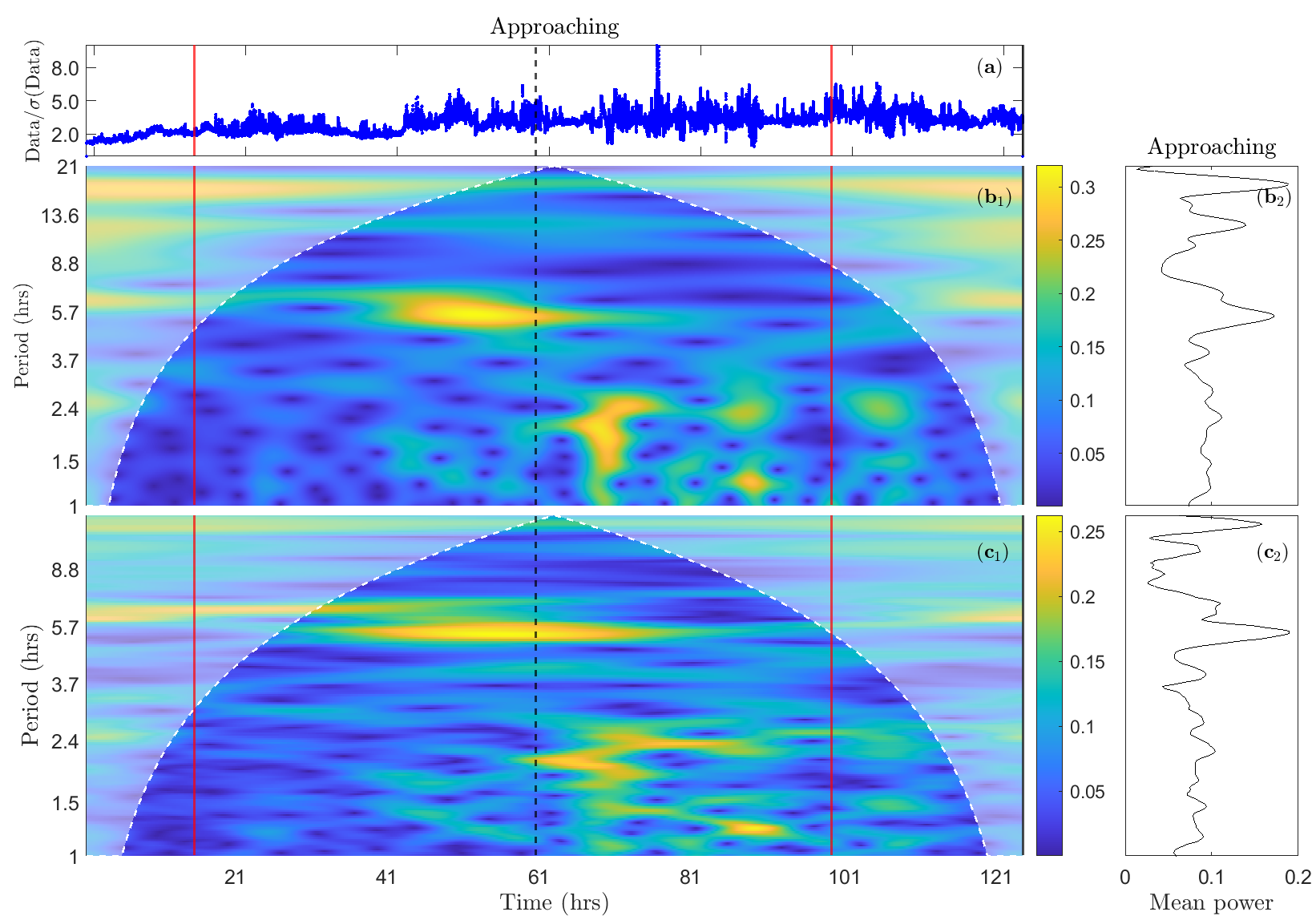}
\caption{Example of CWT analysis of the interpolated normalised fluctuations of density obtained with DMM during the approaching epoch of the Encounter 1 defined within the Procedure (ii). In panel (a), the processed dataset is shown. Panels (b$_1$,b$_2$) and (c$_1$,c$_2$) correspond to Morse and to Bump mother wavelets, respectively. The colour charts (b$_1$,c$_1$) show the dynamical CWT spectra, and the plots (b$_2$, c$_2$) depict their time averages. The faded area in the wavelet spectrum indicates low confidence. The dashed black vertical line denotes the moment when the spacecraft orbital motion is 'purely' radial, and the solid red vertical lines on both sides depict the $\pm 4^{\degree}$ heliocentric longitudinal intervals of the almost radial motion epoch.  \label{Enc1-DMM-A2}}
\end{figure}

\begin{figure}[ht!]%
\includegraphics[scale=0.295]{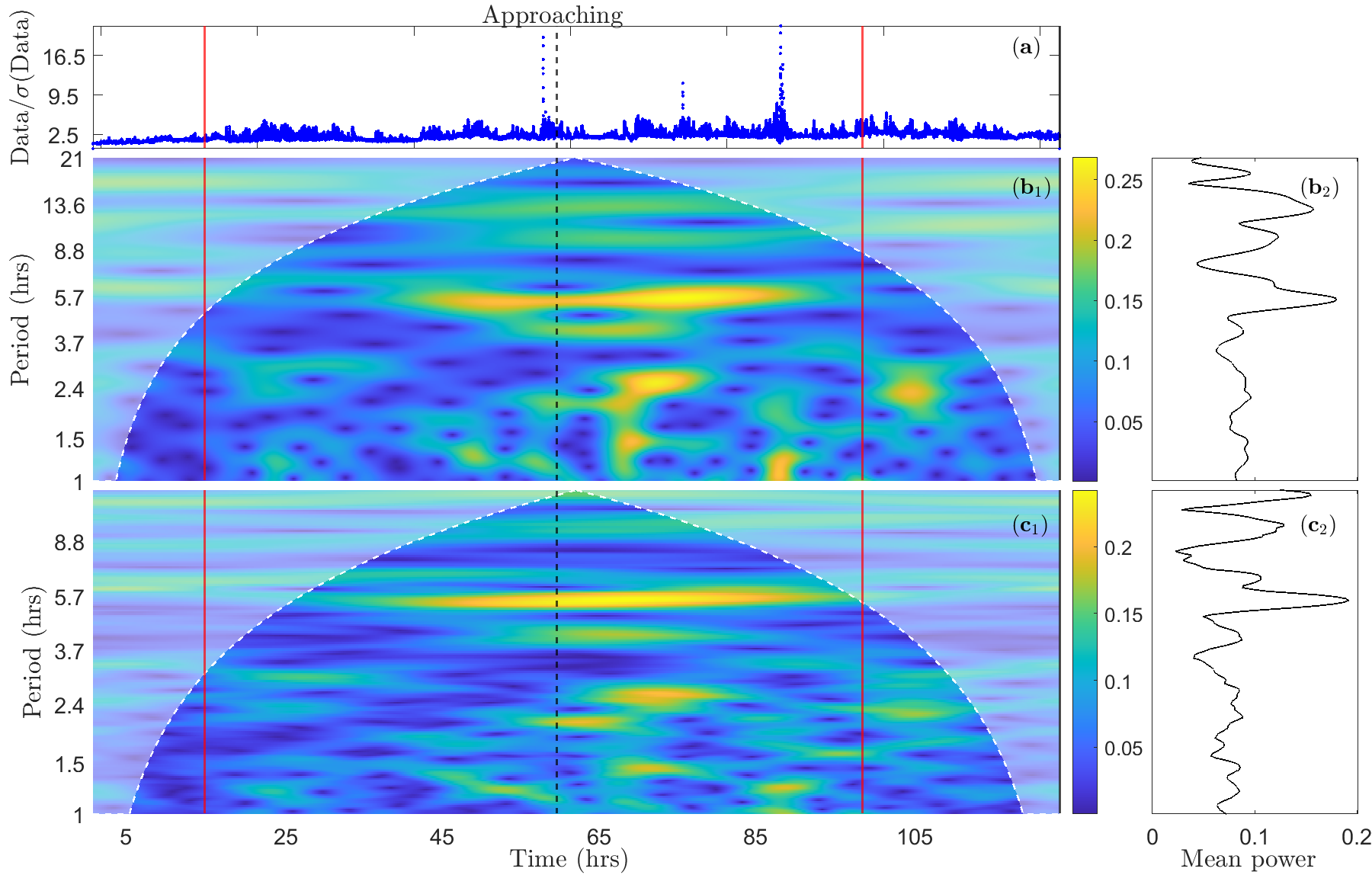}
\caption{Same as in Fig.~\ref{Enc1-DMM-A2}, for the fluctuations of density obtained with FMM. \label{Enc1-FMM-A2}}
\end{figure}

\begin{figure}[ht!]%
\includegraphics[scale=0.32]{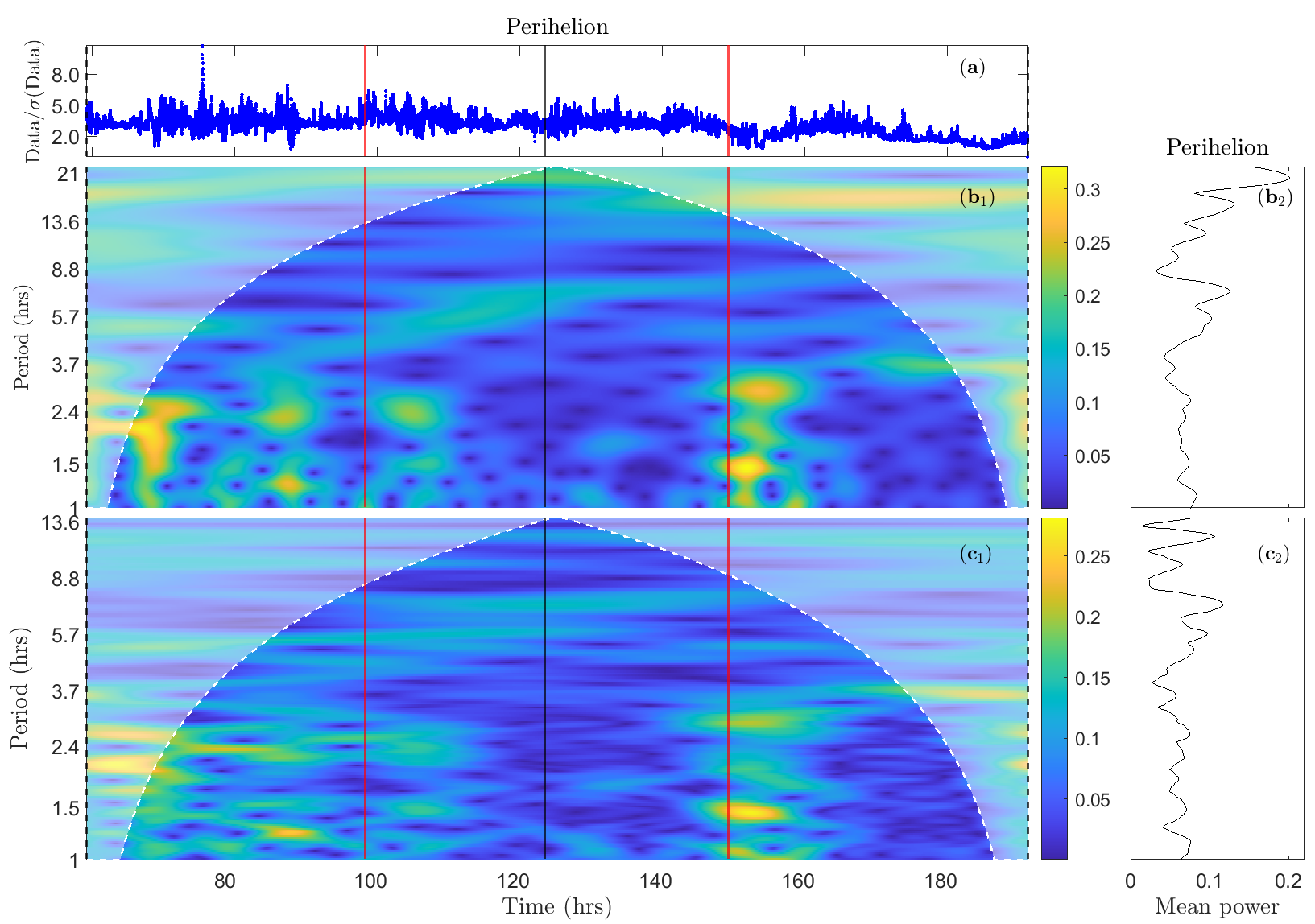}
\caption{Same as in Fig.~\ref{Enc1-DMM-A2}, for perihelion phase of Encounter 1. \label{Enc1-DMM-P2}}
\end{figure}

\begin{figure}[ht!]%
\includegraphics[scale=0.295]{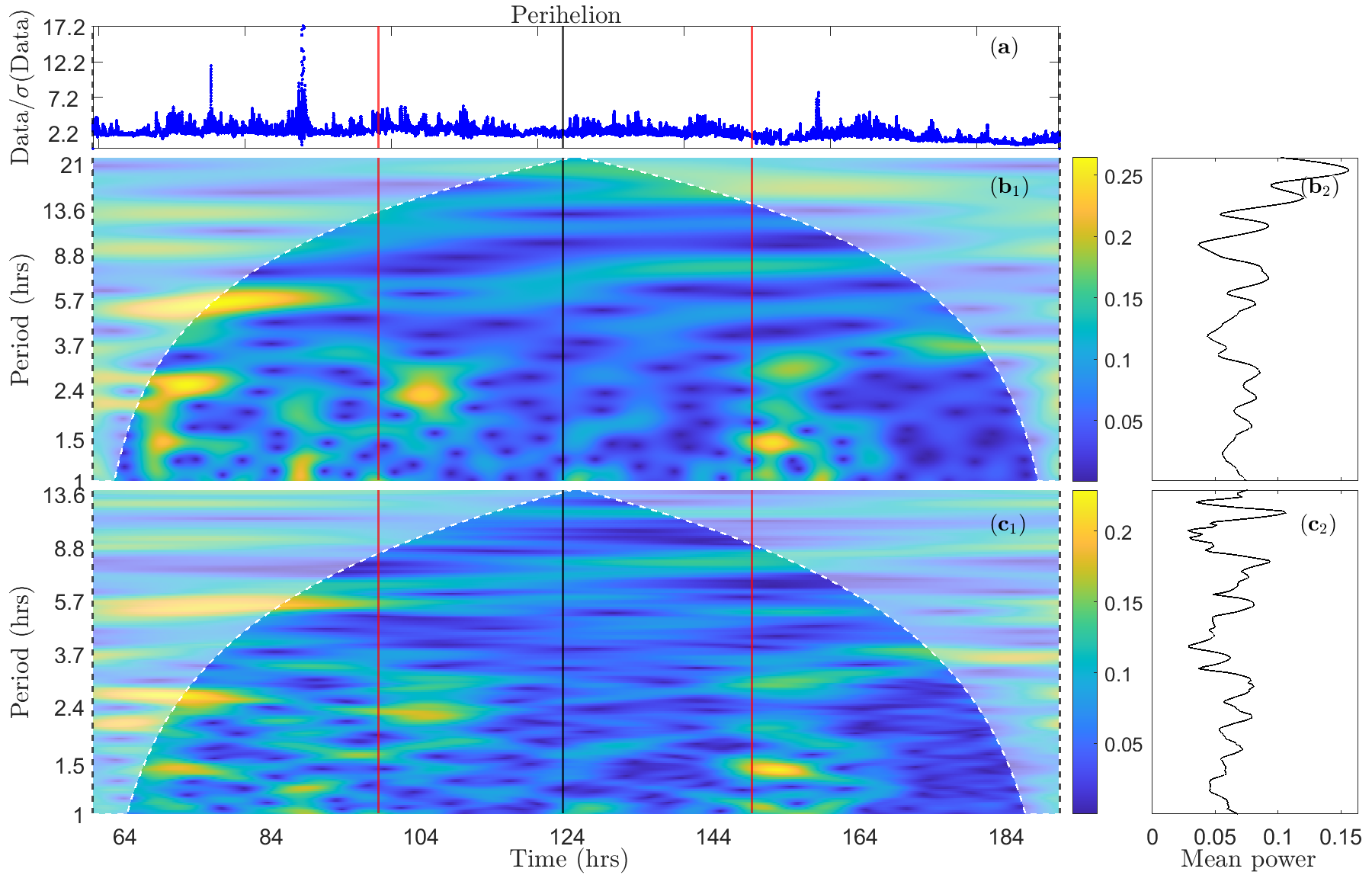}
\caption{Same as in Fig.~\ref{Enc1-DMM-P2}, for the fluctuations of density obtained with FMM.\label{Enc1-FMM-P2}}
\end{figure}

\begin{figure}[ht!]%
\includegraphics[scale=0.325]{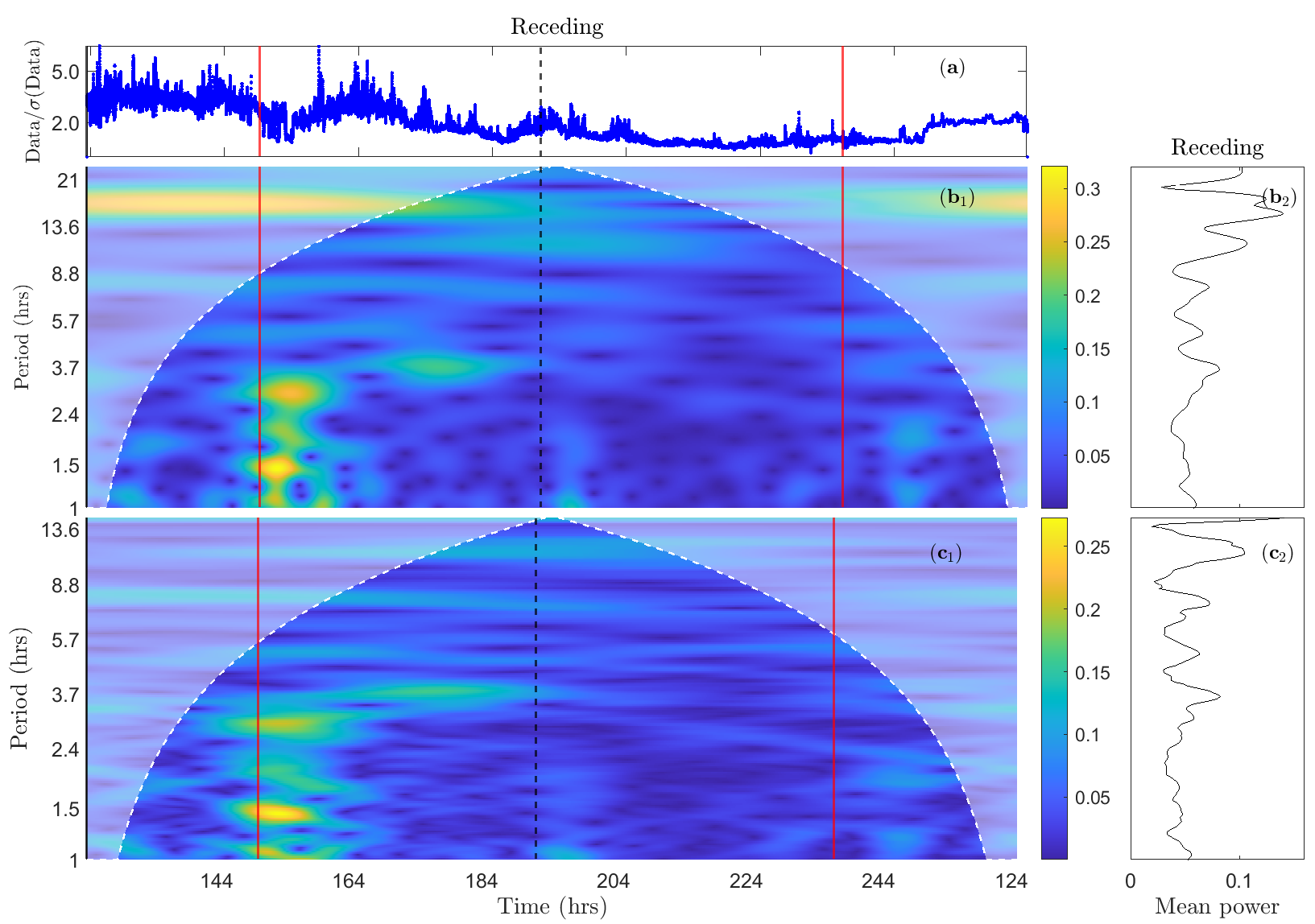}
\caption{Same as in Fig.~\ref{Enc1-DMM-A2}, for receding phase of Encounter 1. \label{Enc1-DMM-R2}}
\end{figure}

\begin{figure}[ht!]%
\includegraphics[scale=0.295]{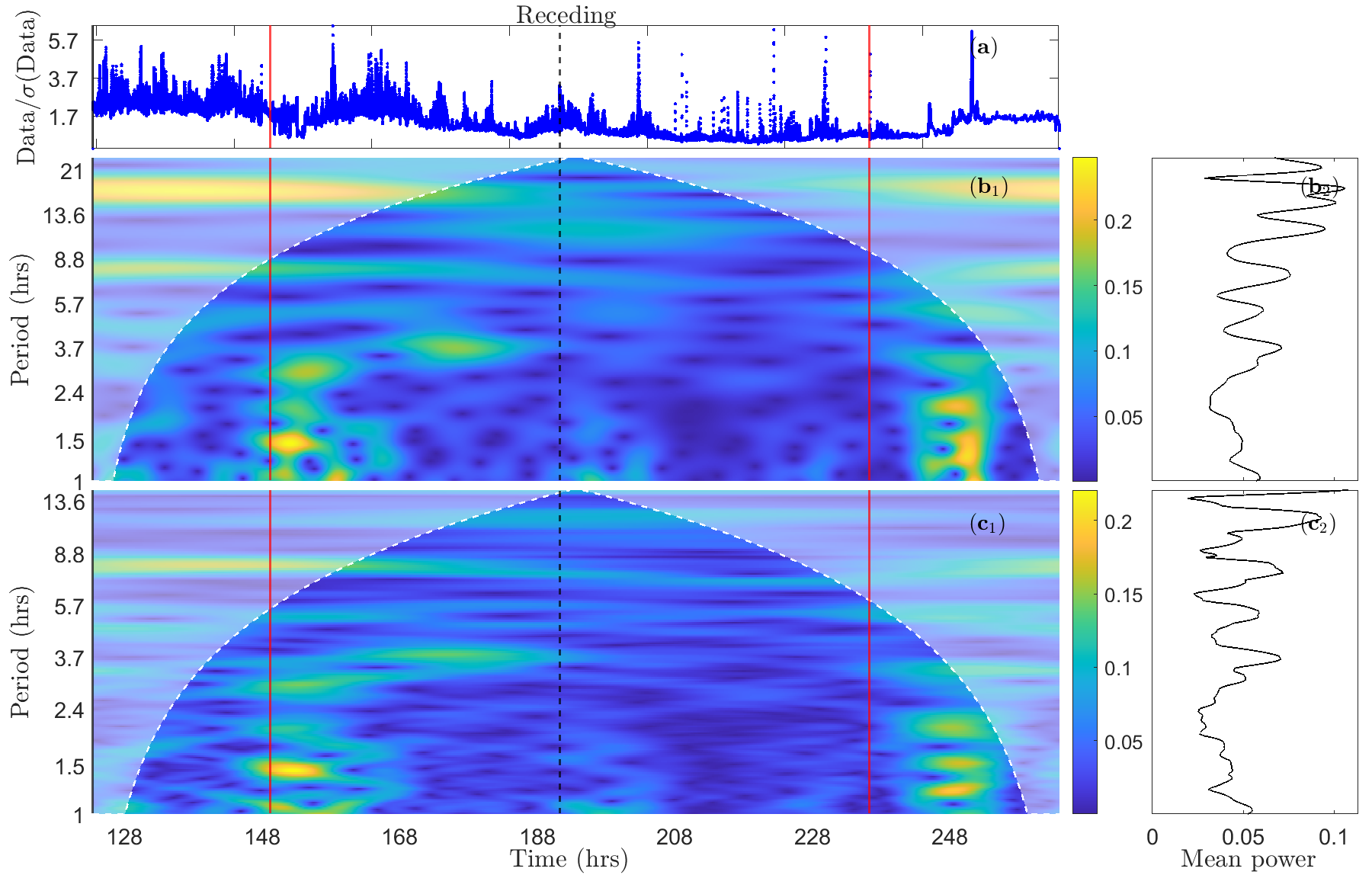}
\caption{ Same as in Fig.~\ref{Enc1-DMM-R2}, for the fluctuations of density obtained with FMM.
\label{Enc1-FMM-R2}}
\end{figure}

\begin{figure}[ht!]%
\includegraphics[scale=0.32]{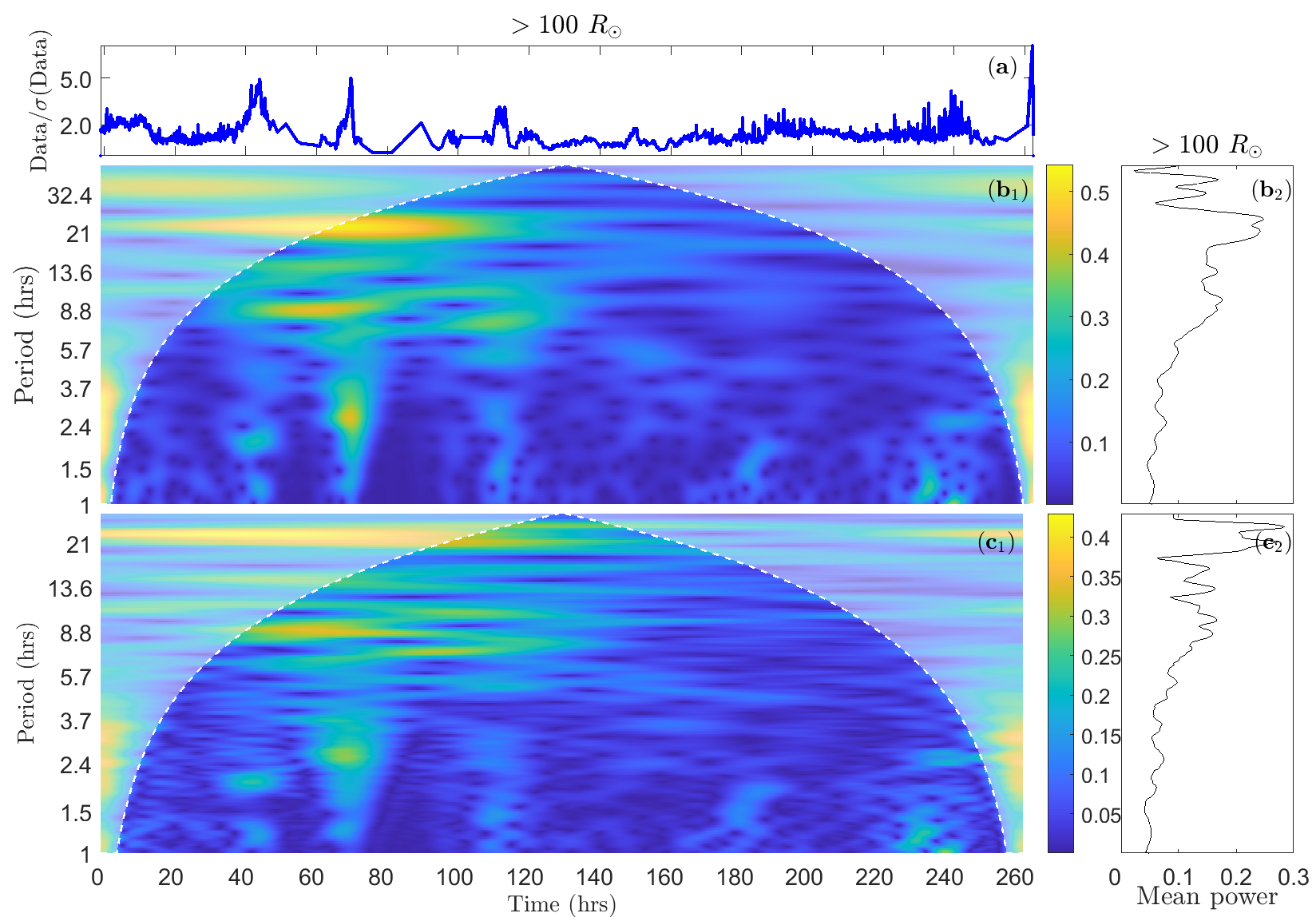}
\caption{Example of CWT analysis of the interpolated normalised fluctuations of density obtained with DMM during the far flight ($>100~R_{\sun}$) epoch after Encounter 1. In panel (a), the processed datasets are shown. Panels (b$_1$,b$_2$) and (c$_1$,c$_2$) correspond to Morse and to Bump mother wavelets, respectively. The colour charts (b$_1$,c$_1$) show the dynamical CWT spectra, and the plots (b$_2$, c$_2$) depict their time averages. The faded area in the wavelet spectrum indicates low confidence. \label{Enc1-DMM-FAR}}
\end{figure}

\begin{figure}[ht!]%
\includegraphics[scale=0.31]{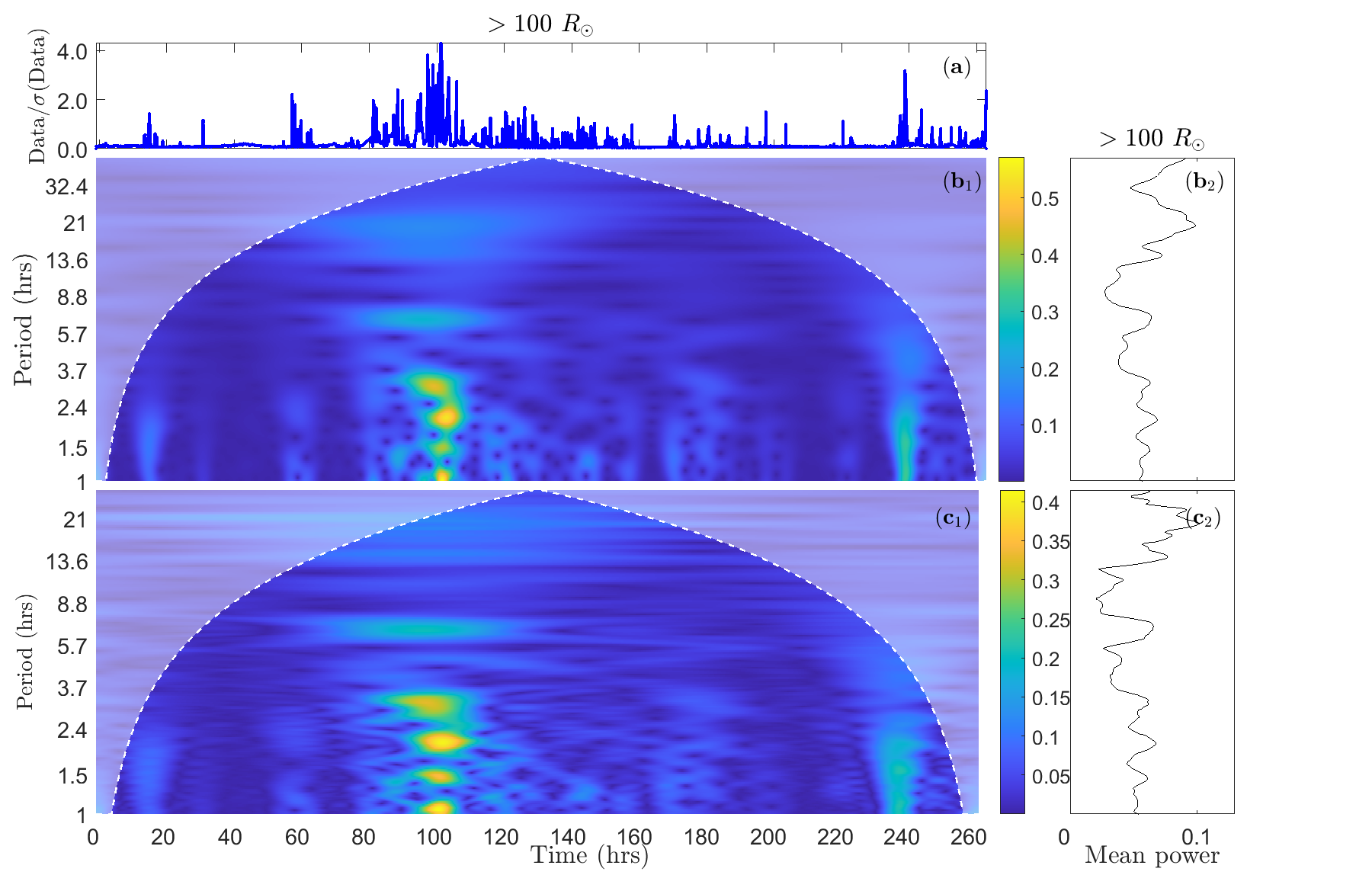}
\caption{Same as in Fig.~\ref{Enc1-DMM-FAR}, for the fluctuations of density obtained with FMM.
\label{Enc1-FMM-FAR}}
\end{figure}

Example results of the performed CWT data analysis of the measured density fluctuations during Encounter 1 are shown in Figs.~\ref{Enc1-DMM-1} - \ref{Enc1-FMM-FAR}, of which Figs.~\ref{Enc1-DMM-1} and \ref{Enc1-FMM-1} present the results obtained with Procedure (i), whereas Figs.~\ref{Enc1-DMM-A2} - \ref{Enc1-FMM-R2} and Figs.~\ref{Enc1-DMM-FAR}, \ref{Enc1-FMM-FAR} correspond to Procedure (ii), and the far flight ($>100~R_{\sun}$) epoch, respectively. The panels (a) in all figures represent the corresponding analysed interpolated datasets with the constant sampling rates normalised by their variances. The indexed panel groups (b) and (c) correspond to CWT spectra obtained with Morse and Bump wavelet functions, respectively, where the colour charts show the dynamical CWT spectra and the plots to the right depict their time averages, which we will refer to further on as global spectra.

Analogous results for all other addressed encounters (Nrs .~2, 4--9, 11--19) are the subject of the following statistical analysis of the detected significant long-periodic oscillations of the considered physical parameters.

Similar to the Lomb-Scargle periodogram analysis, the CWT method reveals several long-periodic modulation features, with typical periods in the range of 3 to 9 hours, as spectrally significant peaks in the average (global) spectra. These peaks are detected independently of the reconstruction method (DMM or FMM) and of the procedure for identifying the encounter epoch. As seen in Figs.~\ref{Enc1-DMM-1}--\ref{Enc1-FMM-R2}, these modulations are well observed during all phases of the encounter (approaching, perihelion, receding). At the same time, the character of detected long-periodic modulations during the far flight epochs (beyond 100 solar radii from the Sun) in Figs.~\ref{Enc1-DMM-FAR}, ~\ref{Enc1-FMM-FAR} is essentially different as compared to the closer encounter flybys, which indicates their different physical nature.

%%%%%%%%%%%%%%%%%%%%%%%%%%%%%%%%%%%%%%%%%%%%%%%%
%%%%%%%%%%%%%%%%%
%%%%%%%%%%%%%%%%%%%%%%%%%%%%%%%%%%%%%%%%%%%%%%%%%%%%%%%%%%%%%%%%%%%%%%%%%%%%%%%%%%%%%%%%%%%
%%%%%%%%%%%
%%%%%%%%%%%%%%%%%%%%%%%%%%%%%%%%%%%%%%%%%%%%%%%%
%%%%%%%%%%%
%%%%%%%%%%%%%%%%%%%%%%%%%%%%%%%%%%%%%%%%%%%%%%%%
%%%%%%%%%%%
%%%%%%%%%%%%%%%%%%%%%%%%%%%%%%%%%%%%%%%%%%%%%%%%
%%%%%%%%%%%
%%%%%%%%%%%%%%%%%%%%%%%%%%%%%%%%%%%%%%%%%%%%%%%%
%%%%%%%%%%%%%%%%%%%%%%%%%%%%%%%%%%%%%%%%%%%%%%%%
\section{Statistical analysis and results}\label{results}
%%%%%%%%%%%%%%%%%%%%%%%%%%%%%%%%%%%%%%%%%%%%%%%%
%%%%%%%%%%%%%%%%%%%%%%%%%%%%%%%%%%%%%%%%%%%%%%%%
\begin{figure*}[ht!]%
\includegraphics[width=1.0\textwidth]{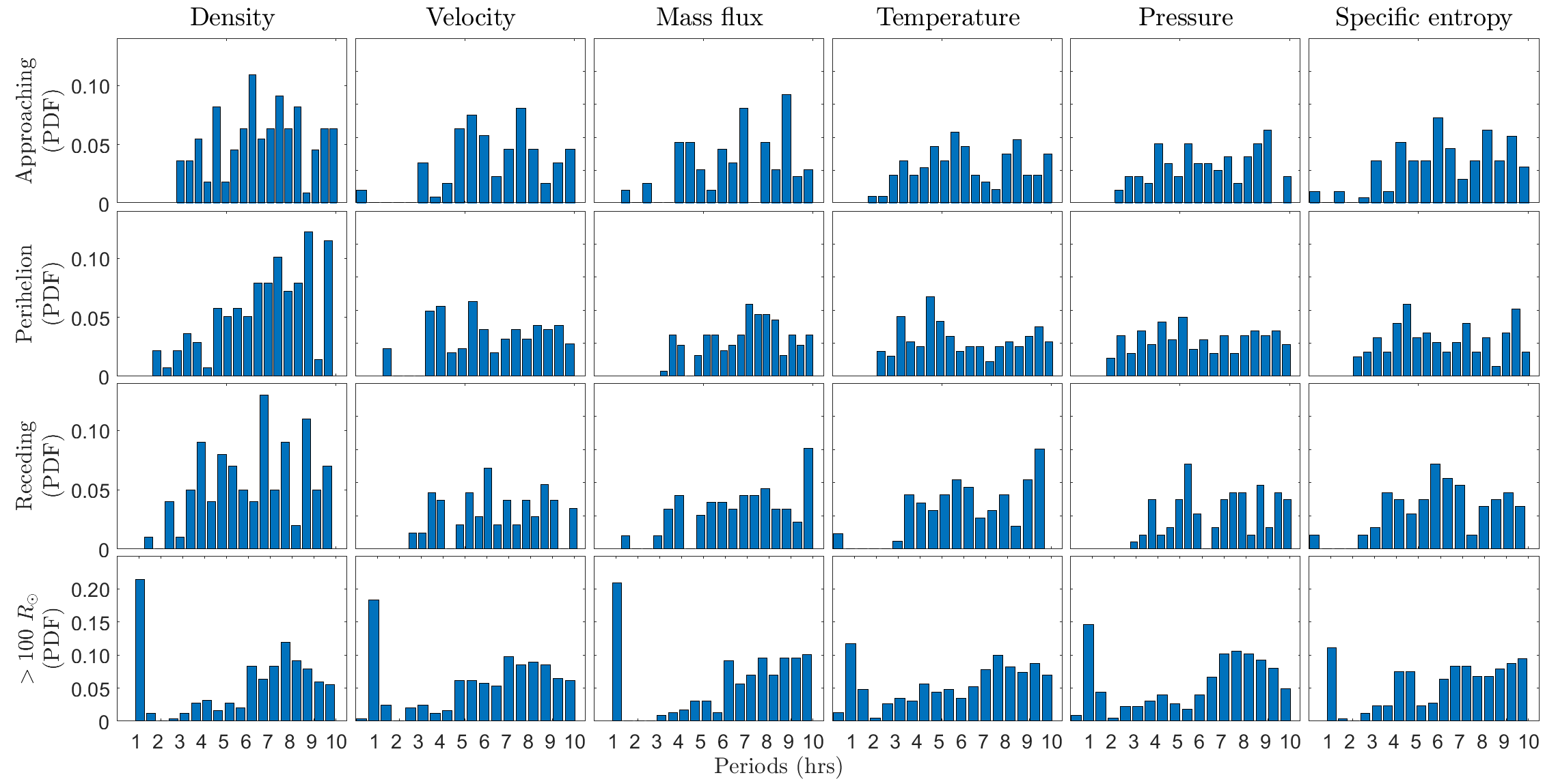}
\caption{Normalized occurrence frequency
histograms (representing Probability Distribution Functions (PDF)) during all considered encounters of the spectrally significant peaks of power in the Lomb-Scargle periodograms for different physical quantities and phases of the flight trajectory obtained from unevenly spaced original datasets normalised by their standard deviations. The width of the bins in each panel is 30 min. 
\label{LS1}}
\end{figure*}
%%%%%%%%%%%%%%%%%%%%%%%%%%%%%%%%%%%%%%%%%%%%%%%%
%%%%%%%%%%%%%%%%%%%%%%%%%%%%%%%%%%%%%%%%%%%%%%%%
\begin{figure}[ht!]%
\centering{\includegraphics[scale=0.34]{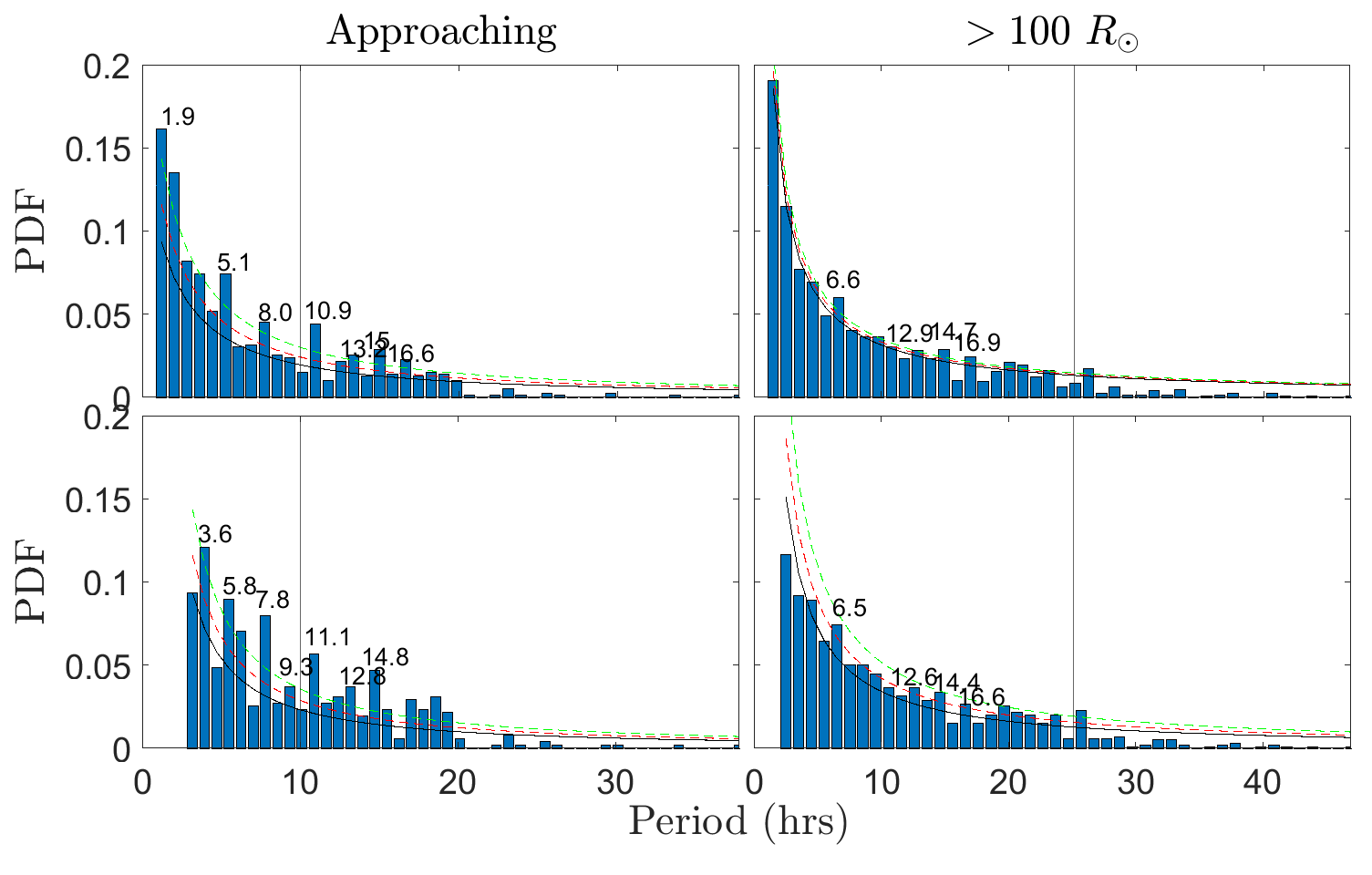}}
\caption{The example of normalized occurrence frequency
histograms (representing Probability Distribution Functions (PDF)) of the power of fluctuations and identifying of the spectrally significant peaks in the CWT mean (global) spectra over the considered period range, during approaching (for the Procedure (i)) and beyond $100~R_{\sun}$ phases of all considered encounters, for the number density observed using both DMM and FMM particle velocity distribution functions and analysed by 'Morse' and 'Bump' mother wavelets. Upper and bottom rows of panels show PDFs for the entire range of periodicities (1--20 hours, option (1)), and the cut period span ( $>3$~hours, option (2)), respectively. The solid blue curves are fitted to the noise power-law function, and the red and green dashed curves represent its $68.2\%$ and $95.4\%$ confidence thresholds.  \label{hist}}
\end{figure}
%%%%%%%%%%%%%%%%%%%%%%%%%%%%%%%%%%%%%%%%%%%%%%%%
As revealed by the Lomb-Scargle and CWT spectral analyses described above, several periodicities were detected in the measured time series of physical quantities, appearing as spectrally significant peaks. Here, we investigate these properties and their statistical significance using both Lomb-Scargle and global wavelet periodograms. The expected result of such an analysis is the selection of those values of characteristic periods for which the high average spectral power in either of the cases could be confirmed with an appropriate statistical confidence. At the same time, our aim is to identify statistically insignificant peaks that can be attributed to statistical noise. To achieve these goals, for each spectral analysis method, we constructed sets of statistical samples of periods from all spectrally significant peaks, classified by different physical quantities and the epochs of the PSP flight trajectory. These sets of peaks have been collected in separate pools, which cover 17 encounters considered in this paper. The next step was to calculate relative occurrence frequencies over the entire range of observed periods, yielding the probability density functions (PDFs) of the revealed periods for each considered case, i.e., for the particular fluctuating physical quantity and specific flight epoch. 
%%%%%%%%%%%%%%%%%%%%%%%%%%%%%%%%%%%%%%%%%%%%%%%%
\subsection{Statistical result 1: Lomb-Scargle power spectra}
%%%%%%%%%%%%%%%%%%%%%%%%%%%%%%%%%%%%%%%%%%%%%%%%
To construct the desired PDFs for Lomb-Scargle periodograms, we have chosen period bins of half-hour width. The obtained PDFs for the physical quantities using the combined pool, which includes estimates based on both the DMM and FMM methods to reconstruct the proton velocity distribution, are shown in Fig.~\ref{LS1}. The panels in Fig.~\ref{LS1} allocated in different columns correspond to physical quantities, whereas the rows from top to bottom show the results for approaching, perihelion, receding and far-flight epochs, respectively. The PDF values in the histograms are computed as the ratio of the spectral peak occurrence rate in each period bin to the total number of peak instances collected from the pool. 

To identify statistically significant peaks in the PDF histograms we applied multi Gaussian fitting procedures for each PDF distribution histograms and evaluated spectral positions of all statistically significant periods (frequencies) and rates of their statistical dispersion around those periods. The later then was used for the calculation of statistical significance expressed by the Cumulative Probability (CP) for each particular period.  In particular, CP is defined as a sum of the PDF bins allocated within the approximating Gaussian function. All PDF bins that lie outside the Gaussian approximations for the identified periodicity peaks are attributed to residual noise, and their sum represents the Residue Cumulative Probability (RCP). In addition, using the obtained values of CP and RCP, we introduced for each histogram Fig.~\ref{LS1}, the coefficient named Relative Statistical Significance (RSS) of the discovered signal, which is evaluated by the following expression. 
\begin{equation}
   \textrm{RSS}=\frac{\sum_{\mathrm{periods}}\textrm{CP}}{\textrm{RCP}}. \label{rss}
\end{equation}
The coefficient RSS is equivalent to the measure of the signal-to-noise ratio.

The complete list of revealed statistically significant periods and corresponding relevant information is provided in Table~\ref{tableLS}. Significant periods, with their uncertainties, are given in the corresponding columns, along with their CP, RCP, and RSS for the particular histogram (i.e., physical quantity and flight epoch). The analysis of the graphs in Fig.~\ref{LS1} immediately shows that the PDFs for the close-by flight trajectories are qualitatively different from those for far-away flight epochs. In most approaches, in the perihelion and receding cases, we have isolated statistically significant periods amid the approximately flat, distributed quasi-white statistical noise with $\textrm{RSS}>1$. For far flight, the noise character is mainly non-flat (power-law), and in most cases $\textrm{RSS}<1$. 
%%%%%%%%%%%%%%%%%%%%%%%%%%%%%%%%%%%%%%%%%%%%%%%%
\subsection{Statistical result 2: Continuous wavelet spectra}\label{CWTstat}
%%%%%%%%%%%%%%%%%%%%%%%%%%%%%%%%%%%%%%%%%%%%%%%%
Similarly to the case of Lomb-Scargle periodograms, we created for each physical quantity over all encounters the pools of the spectrally significant periods related with the local maxima in the CWT average (global) spectra like, e.g., for the number density during encounter one shown in panels (b$_2$)--(b$_4$) and (c$_2$)--(c$_4$) in Figs.~\ref{Enc1-DMM-1}, ~\ref{Enc1-FMM-1}, and (b$_2$), (c$_2$) in Figs.~\ref{Enc1-DMM-A2}--\ref{Enc1-FMM-FAR}. These pools contain all the peaks revealed by our CWT analysis based on both DMM and FMM methods of proton velocity distribution measurements, as well as Morse and Bump mother wavelets over all encounters, still, however, distinguishing between Procedure (i) and Procedure (ii). 

Then, we construct PDF histograms over the corresponding period domains. Examples of such histograms for density fluctuations obtained for the approaching phase within Procedure (i) and for the far flight phase are demonstrated in Fig.~\ref{hist}. The presence of gaps of various widths in the data creates an artificial pollution of the spectra with spurious unphysical features, especially below the level of 3 hours. Therefore, we assume that the obtained CWT spectral range of periodicities of $<3$ hours is strongly affected by the gaps and hence contaminated by unphysical spurious artefacts, which we exclude from consideration in this paper. In view of this, when creating the PDF histograms in Fig.~\ref{hist}, we proceed with two options which consist in consideration of 
\begin{itemize}
    \item Option~1: the entire range of periodicities, i.e., 1--20 hours (upper row of panels in Fig.~\ref{hist}),
    \item Option~2: the cut period span with the periodicities beyond 3 hours only (bottom row of panels in Fig.~\ref{hist}),
\end{itemize}
and compare their outcomes. The number of bins in all histograms is determined by taking the triple square root of the number of period sampling $N_\mathrm{CWT}$ in the averaged global CWT spectra, $N_\mathrm{bins}=3\sqrt{N_\mathrm{CWT}}$. 

The solid blue curves in Fig.~\ref{hist} represent the power-law fits to the distribution. In contrast, red and green dashed curves denote, respectively, the $68.2\%$ and $95.4\%$ confidence levels for the PDF histogram's local peaks. In all cases considered, these lines were defined from the analysis of PDF histograms plotted on a log-log scale. The examples of such log-log plots for option (1) (entire range of 1--20 hour periodicities) and all considered physical quantities over all epochs are shown in Figs.~\ref{CWT1}, ~\ref{CWT2}. 
Note that in these figures, the results obtained using Procedures (i) and (ii) for the close flight epochs (approaching, perihelion, receding) are depicted in the same panels for each physical quantity and artificially shifted from one another to avoid overlaps. Therefore, the $ Y$-Axis values are not shown as conditional. The blue dotted lines (for all considered epochs) indicate the linear fits corresponding to the entire distribution without accounting for any breaking points in the histograms. As can be seen, in most cases, these lines deviate substantially from the actual distribution. Consequently, we assume that the distributions have breaking points within the range $\lg(\mathrm{periods})=1.0-1.3$ (measured in hours) so that the fitted power-law blue solid lines in the diagrams in the upper row of panels in Fig.~\ref{hist} correspond to the data only between the left edge and the break point. Then, similarly, we estimate the deviations of specific data points in the log-log PDF plots relative to the fitted blue solid lines and, assuming that they are normally distributed, calculate the corresponding $\sigma$ and $2\sigma$ levels, shown as red and green dashed lines, respectively. Again, these lines transform into the red and green dashed lines in the PDFs in the upper row of panels of Fig.~\ref{hist}. A similar procedure is performed to calculate the power-law fits for the PDF diagrams in the bottom row of panels in Fig.~\ref{hist} created within option (2), i.e., when considering the cut period span of periodicities ($>3$~hours, up to the breaking point).

The identification of outlier periods within both options (1) and (2) is performed stepwise. First, we identify all peaks in the PDFs, then examine them individually to determine whether they are real detected periods. This is done by comparing the peak levels with the fitted power-law (solid blue line) and its confidence curves (red and green dashed lines). Only those peaks that are at least above the red dashed curve corresponding to $68.2\%$ of confidence are considered statistically relevant. All these peaks are indicated in Figs.~\ref{CWT1}-\ref{CWT2} with red (above $68.2\%$ of confidence level) and green (above $95.4\%$ of confidence level) circles, respectively. Other histogram peaks that appeared below the introduced confidence levels are assumed to be statistically insignificant features and are attributed to statistical noise. We also exclude from the consideration the long-periodic signatures that appeared in the histograms beyond the breaking points in Figs.~\ref{CWT1}-\ref{CWT2}. As can be seen in the CWT spectra in Figs.~\ref{Enc1-DMM-1} and ~\ref{Enc1-FMM-FAR}, the majority of their periods lie outside the confidence area of the spectra. 

A complete list of statistically significant periods, discovered within both options (1) and (2) using the just described method of selection, is consolidated in Tables~\ref{tableCWT1}-\ref{tableCWT2} (for option (1), the entire range of periodicities) and Tables~\ref{tableCWT3}-\ref{tableCWT4} (for option (2), cut range of periodicities, $>3$~hours). In these tables, the numbers shown in brackets, in addition to period values, are the corresponding cumulative probabilities (CP), in analogy to Table~\ref{tableLS} for Lomb-Scargle spectra. Again, we estimate RCP and RSS for each case considered separately. The colours in the table values follow those used to identify outliers.
%%%%%%%%%%%%%%%%%%%%%%%%%%%%%%%%%%%%%%%%%%%%%%%%
\section{Discussion}
%%%%%%%%%%%%%%%%%%%%%%%%%%%%%%%%%%%%%%%%%%%%%%%%
Several immediate outcomes can be inferred from the above-reported statistical analysis of the occurrence rate of spectrally significant oscillations in the considered physical quantities, with periods ranging approximately from 3 to 20 hours. 
%-
In particular, one can observe that the statistics of these oscillations during the close flyby phases of the encounter (approaching, perihelion, and receding) are mutually similar. All of these phases exhibit multiple significant oscillation periods with high statistical confidence. Significant periods are detected most frequently in the range of $< 10$~hours. Such a systematic presence of oscillations in this range is in good agreement with the recently observed characteristic oscillations in active regions (AR) \citep{Dumbadze2021,Dumbadze2017} inferred from SDO/HMI magnetograms \citep{Scherrer12,Schou12}. In particular, statistically significant fluctuation periods of the AR cross-sectional areas and corresponding magnetic fluxes have been detected in the same 3--20 hour range, with the most statistically confident AR periodicities at about 4.5 and 6 hours, which are close to those discussed here. Moreover, in both cases, the solar wind and AR fluctuation spectra show presence of a very similar harmonic discrete structure of the detected statistically significant periods, embedded in a strong background noise making them hardly detectable and requiring special statistical tools for their identification, which we employ in current study. Such a structural similarity in the long-periodic fluctuation spectra of solar wind physical quantities during close flyby phases and in SDO/HMI magnetograms may indicate mutual connection and the presence of common physical background.

In Fig.~\ref{LS1}, one can notice that in the PDFs of the Lomb-Scargle spectra, during the close flyby epochs (top three rows of panels), there are clearly identified statistically significant periodicities that appear mainly above 3 hours, for example, at 4.5 hours, 6--7 hours, etc. In contrast, noticeable periods below 3 hours are absent. At the same time, in the far-flight spectra (bottom row of panels), minor signatures of the 1--3 hour periodicities are clearly seen, being, however, at approximately the same level as the entire fluctuation background. %In view of the fact that the gaps of 1--3 hours duration were removed from the analysed dataset, these hourly features may be related to the real physical processes manifested in the remaining parts of the dataset, especially during the far-flight epoch.

The statistics of all periodicities found with the Lomb-Scargle spectral analysis method are summarised in Table~\ref{tableLS}, from which we can conclude several important aspects:
\begin{itemize}
   \item The statistically significant periods are revealed for the approaching, perihelion and receding phases predominantly in the range of 3--10 hours. The periods attain comparable cumulative occurrence probabilities. 
   \item The detected statistically significant periods during the close flyby epochs are characterised by the RSS essentially higher than 1, indicating a considerable signal-to-noise ratio.
   \item The typical values of several identified statistically significant periodicities appear to be close to those of the AR oscillations. 
   \item The periodic features in the far flight epoch spectra have a noisier character and lower significance with the low values of RSS.
   \item The periodicities below 3 hours in the far flight epoch spectra, despite being visible at the non-flat noisy background, remain statistically relatively insignificant ($RSS<1$), which indicates the weakness and sporadicity of their driving mechanisms.  
\end{itemize}
Regarding the outcome of CWT spectral analysis, the situation with the analysed datasets differs, as the gaps are not literally removed but instead replaced with the linear trends provided by the interpolation procedure. We observed that when considering the entire range of periodicities 1--20 hours (option (1); Tables~\ref{tableCWT1} and \ref{tableCWT2}), rock-solid statistical confidence is given to periods below 3 hours, which at the same time fall to the zone contaminated with possible artefacts related to the linearly fitted gaps in the analysed datasets. Therefore, this type of periodicity should be treated with particular care and cannot be considered as a manifestation of the natural phenomenon. Moreover, exclusion of the hourly part of the periods domain in the case of option (2), as can be seen in Tables~\ref{tableCWT3} and \ref{tableCWT4} makes statistically significant some other, longer-periodic modulations, not appearing so in the case of option (1). At the same time, several periodicities (marked with an asterisk in Tables~\ref{tableCWT1} -- \ref{tableCWT4}) appear statistically significant within both options, and we suppose them to be the most probable existing and related to the real oscillatory process. The comparison of two options for creating the PDF histograms specified in subsection \ref{CWTstat} enables deductions of the following general points: 
\begin{itemize}
   \item The statistically significant periodicities identified within both options are basically similar (i.e., close values), despite some minor deviations in the spectral structure. 
   \item When considering the cut period span (option (2)), the number of revealed significant periods in each case (i.e., physical quantity and flight epoch) is mostly higher as compared to option (1), as the damage imposed on the statistics from the abundant presence of gaps of size up to 3 hours becomes minimised. 
   \item The most trustworthy periodicities revealed within both options again (like in the case of the Lomb-Scargle analysis) fall into the range of typical ARs' oscillations, and show similar discrete spectral structure. 
  \end{itemize}

There is also some insignificant discrepancy between the results obtained with Procedures (i) and (ii), which treat differently the data of encounters and
particular flight epochs (see in subsection \ref{met_wavelet}). It is mainly due to edge effects in the analysed dataset. 

According to results obtained from Lomb-Scargle and wavelet spectral analyses, the appearance of significant periodicities during the far-flung flight epoch ($>100~R_{\sun}$), in contrast to the close flybys, is spurious and lacks statistical confidence. This difference is evident from comparing the estimated CPs of the observed periods with those of related RCPs, which also yield the RSS as a measure of the signal-to-noise ratio. In particular, RSS values presented in Tables~\ref{tableLS} -- \ref{tableCWT4} are systematically much higher than unity for the identified statistically significant periods. On the other hand, the corresponding figures for the far flight epochs in most cases are of the order of one or even less. 

It is also worth noting that the statistically significant periodicities identified by both applied spectral analysis methods are mainly observed in density, bulk radial velocity, corresponding mass flux, and pressure. However, their significance varies across the three phases of the close flyby trajectory. The periods within the same range are also evident in the temperature and specific entropy data, albeit with somewhat reduced statistical power. 

Looking at the periods in Tables~\ref{tableLS}--\ref{tableCWT4}, one can also notice partial differences between the discovered spectral 'harmonics' for different physical quantities. This raises the question of how this is possible for the same plasma, i.e. why different moments of the PDF measured by onboard PSP instruments exhibit different periods? To address this question, it is advantageous to look at the general derivations of particular physical quantities of the solar wind plasma. The instruments on the PSP mission reconstruct the PDFs and calculate their moments, whose properties we examine in the present paper. However, the PDFs can be relatively complex, comprising different particle populations, including suprathermal admixtures, with high anisotropies and significant non-Maxwellian behaviour. This complexity may explain the observed differences in the periodicities of moments of various orders, specifically, density and radial velocity and mass flux on the one hand and temperature, pressure, and entropy on the other. 

In particular, the complex PDF is often composed of more than one particle population, with one being the Maxwellian proton core, superimposed on a non-equilibrium or essentially hotter population of suprathermal particles. The latter contributes mainly to the high-velocity tail of the PDF. In this case, the core of the PDF will primarily determine the behaviour of low-order moments (e.g., density), while the tail dynamics will control higher-order moments. Therefore, if the core population exhibits periodic modulation, this modulation would be most pronounced in the number density. However, if the tail population exhibits periodic modulation, the effect would be less evident in the number density but manifest in higher-order moments, such as temperature. The presence of such complex velocity distributions in the solar wind has recently been reported by \citet{Ofman2025}. This is consistent with our results regarding the slight difference between the observed periodicities of the considered physical quantities.

This concept would mean that such different periods could originate from various particle populations controlled by different physical processes and having different origins. Verifying this hypothesis requires a more thorough and fundamental analysis of the PSP data with a direct investigation of the PDFs, which is beyond the scope of the present work.

%%%%%%%%%%%%%%%%%%%%%%%%%%%%%%%%%%%%%%%%%%%%%%%%
\section{Conclusions} 
%%%%%%%%%%%%%%%%%%%%%%%%%%%%%%%%%%%%%%%%%%%%%%%%
In this work, we investigated datasets of different physical quantities obtained from the SWEAP/PSP instrument during encounters Nrs.~1, 2, 4--9 and 11--19. The particular focus was on checking if similar systematic periodic modulations with proper statistical confidence take place over all phases of the encounter flybys (approaching, perihelion, receding), or they appear only during the perihelion parts of the PSP trajectory \citep{Bale2021}. Another point to consider was the possible link between these periodicities in the inner solar wind and the previously detected global long-periodic oscillations of active regions \citep{Dumbadze2021,Dumbadze2017}. Our key finding is that structure of the spectrum of revealed periods observed in the photospheric magnetic field of active regions and the ambient solar wind are very similar indicating for us that these oscillations have rather non-local, global character that raises the question on availability of the certain physical mechanism standing behind this global link between processes at various levels of solar surrounding. In the current work we just raise the question but not attempt to answer it, which would need an additional analytical or numerical modellings that is beyond the scope of the manuscript. Although, we can state in rather speculative manner on the potential theoretical grounds for the oscillatory phenomena studied here. As the oscillations carry the global nature in the environment of the young solar wind inheriting sequences of periodic wave harmonic patterns, it is plausible to assume that perhaps Alfv\'enic fluctuations confined within the typical solar wind stream pattern are observed. And similarity with the discrete spectra found in ARs may indicate the fact that found oscillations are possibly driven by the photospheric and lower atmospheric processes. One of the candidate mechanism at work can be harmonically modulated turbulent forcing as a consequence of presence of MHD turbulence wave field environment \citep{Pope2000}.
%-
%The specifics of the available SWEAP/PSP data, which we analysed, make it challenging to assess confidently possible modulations with periods less than 3 hours, as this spectral range is strongly affected by data gaps, so that the detected CWT spectral features there may be unphysical artefacts. 
The specifics of the available SWEAP/PSP data, which we analysed, make it challenging to assess possible modulations confidently within periods of less than 3 hours, as this spectral range is strongly affected by data gaps. Hence, the detected CWT spectral features may have unphysical artefacts. 
%-
At the same time, signatures of similar hourly modulations, as revealed by the Lomb-Scargle spectral analysis method, remain relatively weak, indicating a spurious character of the signal. We leave the range of 1--3 hours modulations beyond the scope of our main attention and primarily focus on periodicities of 3--10 hours. We do not speculate here regarding the characteristic spatial scales of the oscillator as it can not be deduced from the single parameter (period) and point measurements we perform in this work. The spatial extent basically depends on the physical nature of the oscillations we see, whether they are Alf\'enic, acoustic or any other type. To make possible judgment on this issue the additional studies are needed, which are currently in progress. 

Consequently, our results lead to several conclusions that are crucially important for understanding the studied coherent, periodic phenomena. It was found that coherent fluctuations represent fundamental properties of the solar surroundings at a global scale, regardless of the spatial orientation (horizontal or radial) of the observing instrument's trajectory. The main result of this work is that several significant, statistically proven periodicities are seen during PSP's almost horizontal perihelion flybys, as well as during the approaching and receding phases with the almost heliocentrically oriented trajectory sections. Our results align and agree with those obtained by white-light images of WISPR/PSP \citep{Poirier2023,Reville2022,Reville2020} as well as with a similar study of the periodic modulations of the inner solar wind background during the perihelion flyby reported in \citet{Bale2021}. 
%-
%In addition, similar periodicities across various physical quantities enable further investigation and the development of theoretical models that elucidate the underlying physical nature of these phenomena. 
%-
In addition, the presence of similar periodicities across various physical quantities enables further investigation and the development of theoretical models that elucidate the underlying physical nature of these phenomena. 
%-
%An additional result is that this significant coherence practically disappears when the PSP measurements are done during the far flights with respect to the encounter region, which leads to the conclusion that the observed coherent phenomena operate predominantly within the solar atmosphere and the inner solar wind (source region). Moreover, the discovered inner solar wind periodicities appear structured similarly as those detected in the active regions \citep{Dumbadze2021,Dumbadze2017}. 
%-
An additional result is that this significant coherence practically disappears when the PSP measurements are conducted during the far flights with respect to the encounter region, leading to the conclusion that the observed coherent phenomena operate predominantly within the solar atmosphere and the inner solar wind (source region). Moreover, the discovered inner solar wind periodicities appear to be structured similarly to those detected in active regions \citep{Dumbadze2021,Dumbadze2017}. 
%-
The signatures of periodicities in the 3--8 hour range are also detectable in the lower solar atmosphere through SDO/AIA data analysis \citep{Wisniewska2024}. Besides, even the periods with ranges below 3 hours are removed from our current analysis; it deserves a separate, accurate attention in the context of the analysis of PSP datasets in the future, as the importance of sub-hourly oscillating features in coronal loops is additionally evidenced recently by the SDO datasets \citep{Zhong2025}.  Such similarity in the long-periodic fluctuation spectra of solar wind physical quantities during the close flyby phases and across various SDO datasets may indicate a possible coupling to a common physical background. While this is a counter-intuitive result, as we know that the solar wind does not originate directly from ARs, it is plausible to assume that magnetic fields of ARs, extended to the corona, and related intermediate magnetic structures may provide such a correlated coherence at the global, solar atmospheric scales, what we observe here. This finding stays in agreement with other indications on available photospheric sources of ambient solar wind \citep[see, e.g.][]{Wang2012}. This raises a question about the physical mechanism that can provide this coupling. In particular, it may indicate the connection between the lower solar atmosphere and the solar wind acceleration region. The mechanism of this connection must become subject to further interest. 

In addition, the presence of relatively weak yet noticeable coherent signatures in temperature and specific entropy datasets indicates the simultaneous presence of different layers of thermodynamically and statistically non-equilibrium plasma with variable temperatures and entropies. Without going into specific theoretical conclusions (beyond the scope of this paper), it can be recalled that the ability of a thermally non-equilibrium plasma to drive various oscillatory modes \citep{shergelashvili07}. Overall, the oscillatory phenomena considered here can be ingredients of stochastically driven far-from-equilibrium statistical processes \citep{Maes_rev2009,Maes2009} in the solar atmosphere and solar wind. In addition, although they fall in different bands of the oscillation spectrum, it is also worth mentioning that the oscillatory processes of periods of a few minutes presumably play a triggering role for the various jet-like flows in the solar atmosphere \citep[see, e.g.][and references therein]{Bagashvili2018}. This reasoning is also supported by the observed discrepancies between the periodicities detected in datasets based on direct measurements of the PDF (DMM) and those based on its Maxwellian fit (FMM). In the first case, the directly measured PDF contains relevant information about the plasma's actual state, including its non-Maxwellian/non-equilibrium components. In contrast, the second approach tends to miss some of this information.

Moreover, the simultaneous presence of different particle populations explains the observed distinction in the periodicities of physical quantities defined by the low- and high-order moments of the PDF, specifically between density, radial velocity, temperature, and entropy. This may be the consequence of distinct oscillatory behaviour in the central and tail parts of the PDF, arising from different physical mechanisms.

Altogether, the results reported in this paper show that common oscillatory phenomena in the solar atmosphere and inner solar wind are a fundamental feature, crucial for understanding the solar wind dynamics. This understanding requires substantial modelling efforts using appropriate analytical and numerical methods (when necessary), extensive observations, and dedicated data analysis. 
%-
\begin{acknowledgements}
The work was supported by a grant from the Shota Rustaveli National Science Foundation of Georgia for fundamental research FR-24-14835, "Phenomenological analysis of solar coronal hole dynamics." SP is funded by the European Union. Views and opinions expressed are, however, those of the author(s) only and do not necessarily reflect those of the European Union or ERCEA. Neither the European Union nor the granting authority can be held responsible for them. This project (Open SESAME) has received funding under the Horizon Europe programme (ERC-AdG agreement No 101141362). These results were also obtained in the framework of the projects C16/24/010 (C1 project Internal Funds KU Leuven), G0B5823N and G002523N (WEAVE) (FWO-Vlaanderen), and 4000145223 (SIDC Data Exploitation (SIDEX2), ESA Prodex). Furthermore, we thankfully acknowledge the financial support for L.W. from the Cusanuswerk via a scholarship programme. We are thankful to Stuart Bale and Nour Rawafi for fruitful discussions during the development of the work. We are thankful to anonymous referee for constructive remarks on our manuscript. 
\end{acknowledgements}

\bibliography{mybib}
%%%%%%%%%%%%%%%%%%%%%%%%%%%%%%%%%%%%%%%%%%%%%%%%
%%%%%%%%%%%%%%%%%%%%%%%%%%%%%%%%%%%%%%%%%%%%%%%%
\begin{appendix}
\section{Tables for the dataset gaps}
  
  \begin{table*}[ht!]
\caption{Total sums of gap durations (only those lasting at least 1 hour are taken in the account) in density datasets measured in hours and the relative percentage of these sums with respect to the total duration of the corresponding flight epoch dataset, at each considered encounter.}
\centering
\begin{tabular}{l c c c c c c}
\hline\hline
Enc. & \multicolumn{3}{c}{DMM} & \multicolumn{3}{c}{FMM} \\
    \hline
 & Approaching & Perihelion & Receding & Approaching & Perihelion & Receding \\
\hline
1  & No gap & No gap & No gap & No gap & No gap & No gap \\
2  & No gap & No gap & 2.5 (2.77 \%) & No gap & No gap & 2.5 (2.77 \%) \\
4  & No gap & 43.0 (47.78 \%) & 6.9 (9.60 \%) & No gap & 44.2 (49.10 \%) & 6.9 (9.60 \%) \\
5  & No gap & 41.8 (43.58 \%) & 36.3 (54.95 \%) & No gap & 41.8 (43.58 \%) & 36.3 (54.95 \%) \\
6  & No gap & 74.7 (79.49 \%) & No gap & No gap & 74.7 (79.49 \%) & No gap \\
7  & No gap & 88.8 (89.38 \%) & 37.1 (65.54 \%) & No gap & 88.8 (89.38 \%) & 37.1 (65.54 \%) \\
8  & 0.9 (1.65 \%) & 82.9 (98.69 \%) & 14.3 (26.51 \%) & 0.9 (1.65 \%) & 82.9 (98.69 \%) & 14.3 (26.51 \%) \\
9  & 1.5 (2.93 \%) & 87.3 (100.00 \%) & 42.9 (75.64 \%) & 1.5 (2.93 \%) & 87.3 (100.00 \%) & 42.9 (75.64 \%) \\
11 & No gap & 64.5 (82.64 \%) & 23.3 (45.93 \%) & No gap & 64.5 (82.64 \%) & 23.3 (45.93 \%) \\
12 & No gap & 67.1 (83.23 \%) & 16.3 (33.95 \%) & No gap & 67.2 (83.25 \%) & 16.3 (33.95 \%) \\
13 & 7.8 (14.38 \%) & 63.1 (77.58 \%) & 2.9 (6.41 \%) & 6.8 (12.67 \%) & 63.1 (77.58 \%) & 2.9 (6.41 \%) \\
14 & No gap & 68.4 (84.13 \%) & No gap & No gap & 68.4 (84.13 \%) & No gap \\
15 & 6.8 (13.39 \%) & 59.8 (73.54 \%) & No gap & 6.8 (13.39 \%) & 59.8 (73.54 \%) & No gap \\
16 & No gap & 73.2 (91.18 \%) & 11.8 (23.19 \%) & No gap & 73.2 (91.18 \%) & 11.8 (23.19 \%) \\
17 & 3.0 (6.82 \%) & 73.6 (97.65 \%) & 9.3 (19.47 \%) & 3.0 (6.82 \%) & 73.6 (97.65 \%) & 9.3 (19.47 \%) \\
18 & 3.4 (7.58 \%) & 75.3 (100.00 \%) & 10.4 (21.70 \%) & 3.4 (7.58 \%) & 75.3 (100.00 \%) & 10.4 (21.70 \%) \\
19 & 11.9 (23.55 \%) & 75.3 (100.00 \%) & 9.6 (21.41 \%) & 11.9 (23.55 \%) & 75.3 (100.00 \%) & 9.6 (21.41 \%) \\
\hline
\end{tabular} 
\label{tableD}
\end{table*}
%%%%%%%%%%%%%%%%%%%%%%%%%%%%%%%%%%%%%%%%%%%%%%%%
%%%%%%%%%%%%%%%%%%%%%%%%%%%%%%%%%%%%%%%%%%%%%%%%

%%%%%------Tables------Velocity
%%%%%%%%%%%%%%%%%%%%%%%%%%%%%%%%%%%%%%%%%%%%%%%%
\begin{table*}[ht!]
\caption{Same as in Table~\ref{tableD} for the bulk radial velocity data.  }
\centering
\begin{tabular}{l c c c c c c}
\hline\hline
Enc. & \multicolumn{3}{c}{DMM} & \multicolumn{3}{c}{FMM} \\
    \hline
 & Approaching & Perihelion & Receding & Approaching & Perihelion & Receding \\
\hline
1  & No gap & No gap & No gap & No gap & No gap & No gap \\
2  & No gap & No gap & 2.5 (2.77 \%) & No gap & No gap & 2.5 (2.77 \%) \\
4  & No gap & 43.0 (47.78 \%) & 6.9 (9.60 \%) & No gap & 44.2 (49.10 \%) & 6.9 (9.60 \%) \\
5  & No gap & 41.8 (43.58 \%) & 36.3 (54.95 \%) & No gap & 41.8 (43.58 \%) & 36.3 (54.95 \%) \\
6  & No gap & 86.6 (92.13 \%) & 14.9 (23.83 \%) & No gap & 86.6 (92.13 \%) & 14.9 (23.83 \%) \\
7  & 1.9 (3.37 \%) & No gap & 37.1 (65.54 \%) & 1.9 (3.37 \%) & No gap & 37.1 (65.54 \%) \\
8  & 20.6 (36.32 \%) & 82.9 (98.69 \%) & 19.3 (35.80 \%) & 20.6 (36.32 \%) & 82.9 (98.69 \%) & 19.3 (35.80 \%) \\
9  & No data & No data & No data & No data & No data & No data \\
11 & No gap & 64.5 (82.64 \%) & 23.3 (45.93 \%) & No gap & 64.5 (82.64 \%) & 23.3 (45.93 \%) \\
12 & No gap & 67.1 (83.23 \%) & 16.3 (33.95 \%) & No gap & 67.2 (83.25 \%) & 16.3 (33.95 \%) \\
13 & 6.8 (12.67 \%) & 63.1 (77.58 \%) & 2.9 (6.41 \%) & 6.8 (12.67 \%) & 63.1 (77.58 \%) & 2.9 (6.41 \%) \\
14 & No gap & 68.4 (84.13 \%) & No gap & No gap & 68.4 (84.13 \%) & No gap \\
15 & 6.8 (13.39 \%) & 59.8 (73.54 \%) & No gap & 6.8 (13.39 \%) & 59.8 (73.54 \%) & No gap \\
16 & No gap & 73.2 (91.18 \%) & 11.8 (23.19 \%) & No gap & 73.2 (91.18 \%) & 11.8 (23.19 \%) \\
17 & 3.0 (6.82 \%) & 73.6 (97.65 \%) & 9.3 (19.47 \%) & 3.0 (6.82 \%) & 73.6 (97.65 \%) & 9.3 (19.47 \%) \\
18 & 3.4 (7.58 \%) & 75.3 (100.00 \%) & 10.4 (21.70 \%) & 3.4 (7.58 \%) & 75.3 (100.00 \%) & 10.4 (21.70 \%) \\
19 & 11.9 (23.55 \%) & 75.3 (100.00 \%) & 9.6 (21.41 \%) & 11.9 (23.55 \%) & 75.3 (100.00 \%) & 9.6 (21.41 \%) \\
\hline
\end{tabular} 
\label{tableV}
\end{table*}
%%%%%%%%%%%%%%%%%%%%%%%%%%%%%%%%%%%%%%%%%%%%%%%%
%%%%%%%%%%%%%%%%%%%%%%%%%%%%%%%%%%%%%%%%%%%%%%%%

%%%%%------Tables------Temperature
%%%%%%%%%%%%%%%%%%%%%%%%%%%%%%%%%%%%%%%%%%%%%%%%
\begin{table*}[ht!]
\caption{Same as in Table~\ref{tableD} for temperature data.  }
\label{tableT}
\centering
\begin{tabular}{l c c c c c c}
\hline\hline
Enc. & \multicolumn{3}{c}{DMM} & \multicolumn{3}{c}{FMM} \\
    \hline
 & Approaching & Perihelion & Receding & Approaching & Perihelion & Receding \\
\hline
1  & No gap & No gap & No gap & No gap & No gap & No gap \\
2  & No gap & No gap & 2.5 (2.77 \%) & No gap & No gap & 2.5 (2.77 \%) \\
4  & No gap & 43.0 (47.78 \%) & 6.9 (9.60 \%) & No gap & 44.2 (49.10 \%) & 6.9 (9.60 \%) \\
5  & No gap & 41.8 (43.58 \%) & 36.3 (54.95 \%) & No gap & 41.8 (43.58 \%) & 36.3 (54.95 \%) \\
6  & No gap & 74.7 (79.49 \%) & No gap & No gap & 74.7 (79.49 \%) & No gap \\
7  & No gap & 88.8 (89.38 \%) & 37.1 (65.54 \%) & No gap & 88.8 (89.38 \%) & 37.1 (65.54 \%) \\
8  & 0.9 (1.65 \%) & 82.9 (98.69 \%) & 14.3 (26.51 \%) & 0.9 (1.65 \%) & 82.9 (98.69 \%) & 14.3 (26.51 \%) \\
9  & 1.5 (2.93 \%) & 87.3 (100.00 \%) & 42.9 (75.64 \%) & 1.5 (2.93 \%) & 87.3 (100.00 \%) & 42.9 (75.64 \%) \\
11 & No gap & 64.5 (82.64 \%) & 23.3 (45.93 \%) & No gap & 64.5 (82.64 \%) & 23.3 (45.93 \%) \\
12 & No gap & 67.1 (83.23 \%) & 16.3 (33.95 \%) & No gap & 67.2 (83.25 \%) & 16.3 (33.95 \%) \\
13 & 6.8 (12.67 \%) & 63.1 (77.58 \%) & 2.9 (6.41 \%) & 6.8 (12.67 \%) & 63.1 (77.58 \%) & 2.9 (6.41 \%) \\
14 & No gap & 68.4 (84.13 \%) & No gap & No gap & 68.4 (84.13 \%) & No gap \\
15 & 6.8 (13.39 \%) & 59.8 (73.54 \%) & No gap & 6.8 (13.39 \%) & 59.8 (73.54 \%) & No gap \\
16 & No gap & 73.2 (91.18 \%) & 11.8 (23.19 \%) & No gap & 73.2 (91.18 \%) & 11.8 (23.19 \%) \\
17 & 3.0 (6.82 \%) & 73.6 (97.65 \%) & 9.3 (19.47 \%) & 3.0 (6.82 \%) & 73.6 (97.65 \%) & 9.3 (19.47 \%) \\
18 & 3.4 (7.58 \%) & 75.3 (100.00 \%) & 10.4 (21.70 \%) & 3.4 (7.58 \%) & 75.3 (100.00 \%) & 10.4 (21.70 \%) \\
19 & 11.9 (23.55 \%) & 75.3 (100.00 \%) & 9.6 (21.41 \%) & 11.9 (23.55 \%) & 75.3 (100.00 \%) & 9.6 (21.41 \%) \\
\hline
\end{tabular} 
\end{table*}
%%%%%%%%%%%%%%%%%%%%%%%%%%%%%%%%%%%%%%%%%%%%%%%%
%%%%%%%%%%%%%%%%%%%%%%%%%%%%%%%%%%%%%%%%%%%%%%%% 

\section{Table demonstrating obtained results using Lomb-Scargle spectral analysis method}
%%%%%%%%%%%%%%%%%%%%%%%%%%%%%%%%%%%%%%%%%%%%%%%%
%%%%%------Tables------
%%%%%%%%%%%%%%%%%%%%%%%%%%%%%%%%%%%%%%%%%%%%%%%%
%\newpage
\begin{table*}[ht!]
\caption{The full set of revealed statistically meaningful periods for different physical quantities and flight epochs obtained using statistics of Lomb-Scargle periodograms analysed in terms of histograms shown in Fig.~\ref{LS1}. the following abbreviations are used: CP -- Cumulative Probability; RCP -- Residue Cumulative Probability denoting CP of the statistical noise; RSS -- Relative Statistical Significance calculated using expression (\ref{rss}). The uncertainty of periods is calculated as half of the width multiplied by a factor of 0.68 (as the single variance of the local Gaussian) of all bins forming the peak of a given period in the histograms. }
\centering
\begin{tabular}{l c c c c c c c c c c c c}
\hline\hline
 & \multicolumn{2}{c}{Density} & \multicolumn{2}{c}{Velocity} & \multicolumn{2}{c}{Mass flux} & \multicolumn{2}{c}{Temperature} & \multicolumn{2}{c}{Pressure} & \multicolumn{2}{c}{Specific entropy} \\
 & Period & CP & Period & CP & Period & CP & Period & CP & Period & CP & Period & CP \\
\hline
\parbox[t]{2mm}{\multirow{5}{*}{\rotatebox[origin=c]{90}{Approaching}}} 
 & 3.3 $\pm$ 0.3 & 0.13 & 3.1 $\pm$ 0.4 & 0.07 & 4.4 $\pm$ 0.3 & 0.24 & 3.2 $\pm$ 0.3 & 0.15 & 2.9 $\pm$ 0.5 & 0.13 & 3.1 $\pm$ 0.4 & 0.09 \\
 & 4.5 $\pm$ 0.3 & 0.12 & 5.3 $\pm$ 0.4 & 0.35 & 5.9 $\pm$ 0.3 & 0.13 & 4.7 $\pm$ 0.3 & 0.2  & 4.2 $\pm$ 0.2 & 0.15 & 4.5 $\pm$ 0.2 & 0.16 \\
 & 6.2 $\pm$ 0.3 & 0.23 & 7.6 $\pm$ 0.4 & 0.31 & 6.8 $\pm$ 0.3 & 0.18 & 5.8 $\pm$ 0.2 & 0.19 & 5.4 $\pm$ 0.3 & 0.19 & 5.9 $\pm$ 0.4 & 0.28 \\
 & 7.2 $\pm$ 0.1 & 0.15 & -- & -- & 7.8 $\pm$ 0.3 & 0.12 & 8.4 $\pm$ 0.3 & 0.22 & 7.2 $\pm$ 0.3 & 0.15 & 7.8 $\pm$ 0.2 & 0.18 \\
 & 8.0 $\pm$ 0.1 & 0.15 & -- & -- & 8.8 $\pm$ 0.3 & 0.23 & -- & -- & 8.5 $\pm$ 0.3 & 0.27 & 9.2 $\pm$ 0.4 & 0.22 \\

RCP & -- & 0.23 & -- & 0.27 & -- & 0.10 & -- & 0.14 & -- & 0.10 & -- & 0.07 \\
RSS & \multicolumn{2}{c}{3.43} & \multicolumn{2}{c}{2.72} & \multicolumn{2}{c}{8.73} & \multicolumn{2}{c}{5.44} & \multicolumn{2}{c}{8.81} & \multicolumn{2}{c}{12.55} \\
\hline
\parbox[t]{2mm}{\multirow{7}{*}{\rotatebox[origin=c]{90}{Perihelion}}}
 & 3.2 $\pm$ 0.3 & 0.09 & 3.7 $\pm$ 0.2 & 0.2  & 3.8 $\pm$ 0.1 & 0.11 & 3.1 $\pm$ 0.3 & 0.17 & 2.3 $\pm$ 0.3 & 0.11 & 3.1 $\pm$ 0.3 & 0.13 \\
 & 4.8 $\pm$ 0.2 & 0.11 & 5.4 $\pm$ 0.3 & 0.23 & 5.3 $\pm$ 0.4 & 0.19 & 4.7 $\pm$ 0.5 & 0.31 & 3.2 $\pm$ 0.3 & 0.11 & 4.2 $\pm$ 0.2 & 0.19 \\
 & 5.7 $\pm$ 0.2 & 0.11 & 7.3 $\pm$ 0.3 & 0.18 & 7.5 $\pm$ 0.5 & 0.44 & 6.5 $\pm$ 0.2 & 0.09 & 4.2 $\pm$ 0.3 & 0.13 & 5.4 $\pm$ 0.3 & 0.17 \\
 & 7.4 $\pm$ 0.3 & 0.25 & 8.8 $\pm$ 0.3 & 0.23 & 9.0 $\pm$ 0.3 & 0.14 & 8.1 $\pm$ 0.3 & 0.14 & 5.1 $\pm$ 0.3 & 0.14 & 7.0 $\pm$ 0.2 & 0.13 \\
 & 8.5 $\pm$ 0.2 & 0.2  & -- & -- & -- & -- & 9.4 $\pm$ 0.3 & 0.19 & 6.1 $\pm$ 0.3 & 0.09 & 7.9 $\pm$ 0.2 & 0.09 \\
 & -- & -- & -- & -- & -- & -- & -- & -- & 7.0 $\pm$ 0.3 & 0.11 & 9.2 $\pm$ 0.2 & 0.17 \\
 & -- & -- & -- & -- & -- & -- & -- & -- & 8.7 $\pm$ 0.5 & 0.26 & -- & -- \\

RCP & -- & 0.24 & -- & 0.16 & -- & 0.12 & -- & 0.10 & -- & 0.05 & -- & 0.12 \\
RSS & \multicolumn{2}{c}{3.11} & \multicolumn{2}{c}{5.19} & \multicolumn{2}{c}{7.57} & \multicolumn{2}{c}{9.21} & \multicolumn{2}{c}{19.81} & \multicolumn{2}{c}{7.59} \\
\hline
\parbox[t]{2mm}{\multirow{6}{*}{\rotatebox[origin=c]{90}{Receding}}}
 & 3.8 $\pm$ 0.3 & 0.18 & 3.5 $\pm$ 0.3 & 0.18 & 3.9 $\pm$ 0.3 & 0.14 & 3.8 $\pm$ 0.6 & 0.22 & 3.7 $\pm$ 0.3 & 0.12 & 3.9 $\pm$ 0.6 & 0.24 \\
 & 5.0 $\pm$ 0.2 & 0.15 & 5.0 $\pm$ 0.1 & 0.12 & 5.6 $\pm$ 0.5 & 0.25 & 5.7 $\pm$ 0.4 & 0.28 & 5.4 $\pm$ 0.3 & 0.26 & 6.1 $\pm$ 0.6 & 0.4  \\
 & 6.7 $\pm$ 0.3 & 0.22 & 5.8 $\pm$ 0.1 & 0.17 & 7.6 $\pm$ 0.5 & 0.31 & 7.6 $\pm$ 0.6 & 0.22 & 7.4 $\pm$ 0.3 & 0.24 & 8.8 $\pm$ 0.6 & 0.29 \\
 & 7.9 $\pm$ 0.2 & 0.11 & 6.7 $\pm$ 0.1 & 0.11 & -- & -- & -- & -- & 8.6 $\pm$ 0.3 & 0.15 & -- & -- \\
 & 8.9 $\pm$ 0.2 & 0.16 & 7.8 $\pm$ 0.3 & 0.16 & -- & -- & -- & -- & 9.7 $\pm$ 0.1 & 0.16 & -- & -- \\
 & -- & -- & 8.9 $\pm$ 0.1 & 0.17 & -- & -- & -- & -- & -- & -- & -- & -- \\

RCP & -- & 0.18 & -- & 0.09 & -- & 0.29 & -- & 0.28 & -- & 0.07 & -- & 0.06 \\
RSS & \multicolumn{2}{c}{4.56} & \multicolumn{2}{c}{10.66} & \multicolumn{2}{c}{2.39} & \multicolumn{2}{c}{2.58} & \multicolumn{2}{c}{12.49} & \multicolumn{2}{c}{14.57} \\
\hline
RSS & \multicolumn{2}{c}{$<$0.35} & \multicolumn{2}{c}{$<$0.37} & \multicolumn{2}{c}{$<$0.28} & \multicolumn{2}{c}{$<$0.48} & \multicolumn{2}{c}{$<$0.17} & \multicolumn{2}{c}{$<$0.19} \\
\hline
\end{tabular} 
\label{tableLS}
\end{table*}
%%%%%%%%%%%%%%%%%%%%%%%%%%%%%%%%%%%%%%%%%%%%%%%%
%%%%%%%%%%%%%%%%%%%%%%%%%%%%%%%%%%%%%%%%%%%%%%%%
\section{Tables and plots demonstrating obtained results using CWT spectral analysis method}
%%%%%%%%%%%%%%%%%%%%%%%%%%%%%%%%%%%%%%%%%%%%%%%%
\newpage
\begin{table*}[ht!]
\caption{Same as in Table~\ref{tableLS} for statistics of significant periods shown by coloured fields in the table and measured in hours for density, bulk radial velocity and mass flux, obtained using plots like in Fig.~\ref{CWT1} and the PDF histograms like in the upper row of Fig.~\ref{hist}, i.e. for the option (1) (entire range of periodicities, 1--20 hours). The notation of the statistical parameters is the same as in Table~\ref{tableLS} (the numbers shown in brackets, aside each period value, represent corresponding CPs). The colouring of the revealed periods follows the highlights of significant points, with circles in Fig.~\ref{CWT1}. }
\label{tableCWT1}
\centering
\begin{tabular}{l c c c c c c}
\hline\hline
 & \multicolumn{2}{c}{Density} & \multicolumn{2}{c}{Velocity} & \multicolumn{2}{c}{Mass Flux} \\
 & Procedure 1 & Procedure 2 & Procedure 1 & Procedure 2 & Procedure 1 & Procedure 2 \\
\hline
\parbox[t]{2mm}{\multirow{7}{*}{\rotatebox[origin=c]{90}{Approaching}}} 
 & \color[HTML]{008000}{5.1 $\pm$ 0.6 (0.19)} & \color[HTML]{FF0000}{1.5 $\pm$ 0.3 (0.31)} & \color[HTML]{008000}{1.7 $\pm$ 0.5 (0.54)} & \color[HTML]{008000}{1.5 $\pm$ 0.3 (0.42)} & \color[HTML]{008000}{4.1 $\pm$ 0.4 (0.20)} & \color[HTML]{FF0000}{2.9 $\pm$ 0.3 (0.11)} \\
 & \color[HTML]{008000}{8.0$^*$ $\pm$ 0.8 (0.09)} & \color[HTML]{FF0000}{2.8 $\pm$ 0.3 (0.11)} & \color[HTML]{008000}{3.6 $\pm$ 0.5 (0.14)} & \color[HTML]{FF0000}{2.7 $\pm$ 0.3 (0.15)} & \color[HTML]{008000}{5.8$^*$ $\pm$ 0.4 (0.17)} & \color[HTML]{FF0000}{3.9 $\pm$ 0.3 (0.11)} \\
 & \color[HTML]{008000}{10.9$^*$ $\pm$ 0.6 (0.19)} & \color[HTML]{008000}{4.1 $\pm$ 0.3 (0.14)} & \color[HTML]{008000}{5.3$^*$ $\pm$ 0.7 (0.09)} & \color[HTML]{FF0000}{3.9$^*$ $\pm$ 0.3 (0.05)} & \color[HTML]{008000}{8.0$^*$ $\pm$ 0.8 (0.09)} & \color[HTML]{FF0000}{5.3 $\pm$ 0.3 (0.15)} \\
 & \color[HTML]{008000}{13.2 $\pm$ 0.6 (0.07)} & \color[HTML]{008000}{5.0$^*$ $\pm$ 0.3 (0.10)} & \color[HTML]{008000}{7.8$^*$ $\pm$ 0.5 (0.08)} & \color[HTML]{FF0000}{4.8$^*$ $\pm$ 0.3 (0.07)} & \color[HTML]{008000}{10.2 $\pm$ 0.4 (0.11)} & \color[HTML]{FF0000}{6.8$^*$ $\pm$ 0.3 (0.08)} \\
 & \color[HTML]{008000}{15.0$^*$ $\pm$ 0.6 (0.08)} & \color[HTML]{008000}{6.3 $\pm$ 0.3 (0.14)} & \color[HTML]{008000}{10.2$^*$ $\pm$ 0.7 (0.05)} & \color[HTML]{FF0000}{6.0$^*$ $\pm$ 0.3 (0.08)} & \color[HTML]{008000}{11.6 $\pm$ 0.4 (0.27)} & \color[HTML]{FF0000}{7.8$^*$ $\pm$ 0.3 (0.06)} \\
 & \color[HTML]{008000}{16.6 $\pm$ 0.6 (0.05)} & \color[HTML]{008000}{8.0$^*$ $\pm$ 0.3 (0.14)} & \color[HTML]{008000}{15.4$^*$ $\pm$ 0.5 (0.07)} & \color[HTML]{008000}{7.8$^*$ $\pm$ 0.4 (0.11)} & -- & \color[HTML]{FF0000}{8.7$^*$ $\pm$ 0.3 (0.06)} \\
 & -- & -- & -- & \color[HTML]{FF0000}{9.1$^*$ $\pm$ 0.3 (0.09)} & -- & \color[HTML]{FF0000}{10.3$^*$ $\pm$ 0.3 (0.08)} \\

RCP & 0.34 & 0.06 & 0.03 & 0.02 & 0.16 & 0.35 \\
RSS & 1.95 & 16.43 & 39.00 & 42.98 & 5.26 & 1.87 \\
\hline
\parbox[t]{2mm}{\multirow{9}{*}{\rotatebox[origin=c]{90}{Perihelion}}}
 & \color[HTML]{FF0000}{6.4 $\pm$ 1.0 (0.20)} & \color[HTML]{FF0000}{2.8 $\pm$ 0.4 (0.19)} & \color[HTML]{FF0000}{2.2 $\pm$ 0.8 (0.31)} & \color[HTML]{008000}{1.5 $\pm$ 0.3 (0.25)} & \color[HTML]{008000}{2.2 $\pm$ 0.9 (0.41)} & \color[HTML]{008000}{2.8 $\pm$ 0.4 (0.23)} \\
 & \color[HTML]{008000}{9.7$^*$ $\pm$ 0.6 (0.11)} & \color[HTML]{FF0000}{4.4$^*$ $\pm$ 0.4 (0.08)} & \color[HTML]{FF0000}{5.6$^*$ $\pm$ 0.8 (0.12)} & \color[HTML]{FF0000}{2.5 $\pm$ 0.3 (0.11)} & \color[HTML]{008000}{5.3$^*$ $\pm$ 0.4 (0.09)} & \color[HTML]{008000}{4.3$^*$ $\pm$ 0.2 (0.11)} \\
 & \color[HTML]{008000}{11.6$^*$ $\pm$ 0.6 (0.13)} & \color[HTML]{FF0000}{5.6$^*$ $\pm$ 0.4 (0.07)} & \color[HTML]{FF0000}{7.5$^*$ $\pm$ 0.6 (0.08)} & \color[HTML]{FF0000}{4.8$^*$ $\pm$ 0.5 (0.08)} & \color[HTML]{FF0000}{6.5$^*$ $\pm$ 0.4 (0.08)} & \color[HTML]{008000}{5.4$^*$ $\pm$ 0.2 (0.09)} \\
 & \color[HTML]{008000}{14.5$^*$ $\pm$ 0.6 (0.14)} & \color[HTML]{008000}{6.9$^*$ $\pm$ 0.6 (0.14)} & \color[HTML]{008000}{10.4$^*$ $\pm$ 0.8 (0.13)} & \color[HTML]{008000}{7.0$^*$ $\pm$ 0.3 (0.08)} & \color[HTML]{FF0000}{7.8$^*$ $\pm$ 0.4 (0.06)} & \color[HTML]{008000}{6.6$^*$ $\pm$ 0.4 (0.16)} \\
 & \color[HTML]{008000}{16.8 $\pm$ 1.0 (0.10)} & \color[HTML]{008000}{10.0$^*$ $\pm$ 0.4 (0.11)} & \color[HTML]{008000}{15.6$^*$ $\pm$ 1.1 (0.21)} & \color[HTML]{FF0000}{7.9$^*$ $\pm$ 0.3 (0.04)} & \color[HTML]{FF0000}{9.2 $\pm$ 0.4 (0.04)} & \color[HTML]{008000}{8.0$^*$ $\pm$ 0.2 (0.07)} \\
 & -- & \color[HTML]{008000}{11.6 $\pm$ 0.4 (0.11)} & \color[HTML]{008000}{19.1$^*$ $\pm$ 0.6 (0.11)} & \color[HTML]{008000}{9.2$^*$ $\pm$ 0.3 (0.07)} & \color[HTML]{FF0000}{10.9 $\pm$ 0.4 (0.02)} & \color[HTML]{008000}{9.8 $\pm$ 0.4 (0.06)} \\
 & -- & \color[HTML]{FF0000}{13.3 $\pm$ 0.4 (0.03)} & -- & \color[HTML]{008000}{10.4$^*$ $\pm$ 0.3 (0.11)} & \color[HTML]{FF0000}{12.9 $\pm$ 0.9 (0.11)} & \color[HTML]{008000}{11.4 $\pm$ 0.4 (0.15)} \\
 & -- & \color[HTML]{008000}{14.1 $\pm$ 0.4 (0.04)} & -- & \color[HTML]{008000}{12.7$^*$ $\pm$ 0.3 (0.11)} & \color[HTML]{008000}{17.3 $\pm$ 0.9 (0.14)} & -- \\
 & -- & \color[HTML]{FF0000}{16.3 $\pm$ 0.6 (0.03)} & -- & \color[HTML]{008000}{15.2$^*$ $\pm$ 0.3 (0.08)} & -- & -- \\

RCP & 0.31 & 0.19 & 0.04 & 0.08 & 0.05 & 0.17 \\
RSS & 2.18 & 4.40 & 25.72 & 12.04 & 21.15 & 4.96 \\
\hline
\parbox[t]{2mm}{\multirow{8}{*}{\rotatebox[origin=c]{90}{Receding}}}
 & \color[HTML]{FF0000}{2.0 $\pm$ 0.7 (0.63)} & \color[HTML]{FF0000}{5.4$^*$ $\pm$ 0.4 (0.12)} & \color[HTML]{008000}{1.6 $\pm$ 0.5 (0.39)} & \color[HTML]{FF0000}{2.7 $\pm$ 0.4 (0.14)} & \color[HTML]{FF0000}{2.5 $\pm$ 0.5 (0.18)} & \color[HTML]{008000}{5.6$^*$ $\pm$ 0.4 (0.20)} \\
 & \color[HTML]{008000}{5.9 $\pm$ 0.5 (0.11)} & \color[HTML]{008000}{7.0$^*$ $\pm$ 0.4 (0.09)} & \color[HTML]{008000}{3.5 $\pm$ 0.5 (0.15)} & \color[HTML]{008000}{3.8$^*$ $\pm$ 0.4 (0.15)} & \color[HTML]{FF0000}{4.6$^*$ $\pm$ 0.5 (0.15)} & \color[HTML]{008000}{7.3$^*$ $\pm$ 0.5 (0.18)} \\
 & \color[HTML]{FF0000}{8.0$^*$ $\pm$ 0.5 (0.04)} & \color[HTML]{008000}{7.9 $\pm$ 0.4 (0.19)} & \color[HTML]{008000}{5.0 $\pm$ 0.5 (0.08)} & \color[HTML]{008000}{5.3$^*$ $\pm$ 0.4 (0.15)} & \color[HTML]{FF0000}{7.9$^*$ $\pm$ 0.7 (0.13)} & \color[HTML]{FF0000}{9.7$^*$ $\pm$ 0.4 (0.11)} \\
 & \color[HTML]{FF0000}{10.1$^*$ $\pm$ 0.5 (0.05)} & \color[HTML]{008000}{10.1$^*$ $\pm$ 0.7 (0.23)} & \color[HTML]{008000}{7.3$^*$ $\pm$ 0.7 (0.12)} & \color[HTML]{008000}{7.4$^*$ $\pm$ 0.4 (0.14)} & \color[HTML]{FF0000}{10.9$^*$ $\pm$ 0.7 (0.13)} & \color[HTML]{FF0000}{12.3$^*$ $\pm$ 0.4 (0.06)} \\
 & -- & -- & \color[HTML]{008000}{9.6$^*$ $\pm$ 0.5 (0.06)} & \color[HTML]{008000}{8.9$^*$ $\pm$ 0.4 (0.11)} & \color[HTML]{FF0000}{14.3 $\pm$ 0.5 (0.06)} & \color[HTML]{008000}{13.4 $\pm$ 0.4 (0.12)} \\
 & -- & -- & \color[HTML]{FF0000}{11.7$^*$ $\pm$ 0.5 (0.04)} & \color[HTML]{008000}{10.1$^*$ $\pm$ 0.5 (0.13)} & \color[HTML]{FF0000}{15.6$^*$ $\pm$ 0.5 (0.04)} & -- \\
 & -- & -- & \color[HTML]{008000}{13.3$^*$ $\pm$ 0.7 (0.10)} & -- & \color[HTML]{FF0000}{18.6 $\pm$ 0.5 (0.05)} & -- \\
 & -- & -- & \color[HTML]{FF0000}{17.1$^*$ $\pm$ 0.5 (0.03)} & -- & -- & -- \\

RCP & 0.16 & 0.36 & 0.03 & 0.18 & 0.26 & 0.33 \\
RSS & 5.11 & 1.75 & 29.00 & 4.50 & 2.87 & 2.05 \\
\hline
\parbox[t]{2mm}{\multirow{4}{*}{\rotatebox[origin=c]{90}{$>100~R_{\sun}$}}}
 & \multicolumn{2}{c}{\color[HTML]{FF0000}{6.6$^*$ $\pm$ 0.7 (0.22)} } & \multicolumn{2}{c}{\color[HTML]{008000}{4.8$^*$ $\pm$ 0.3 (0.19)} } & \multicolumn{2}{c}{\color[HTML]{FF0000}{6.6$^*$ $\pm$ 0.7 (0.23)} } \\
 & \multicolumn{2}{c}{\color[HTML]{FF0000}{12.8$^*$ $\pm$ 0.7 (0.08)} } & \multicolumn{2}{c}{\color[HTML]{FF0000}{7.1 $\pm$ 0.7 (0.11)} } & \multicolumn{2}{c}{\color[HTML]{008000}{9.7$^*$ $\pm$ 0.7 (0.13)} } \\
 & \multicolumn{2}{c}{\color[HTML]{008000}{14.7$^*$ $\pm$ 0.7 (0.15)} } & \multicolumn{2}{c}{\color[HTML]{008000}{11.4 $\pm$ 0.3 (0.14)} } & \multicolumn{2}{c}{\color[HTML]{008000}{14.7$^*$ $\pm$ 0.7 (0.13)}} \\
 & \multicolumn{2}{c}{\color[HTML]{008000}{16.9$^*$ $\pm$ 0.7 (0.22)} } & \multicolumn{2}{c}{\color[HTML]{008000}{13.4 $\pm$ 0.3 (0.07)} } & \multicolumn{2}{c}{\color[HTML]{008000}{17.0 $\pm$ 0.7 (0.14)} } \\
 & \multicolumn{2}{c}{--} & \multicolumn{2}{c}{\color[HTML]{008000}{16.2 $\pm$ 0.3 (0.11)} } & \multicolumn{2}{c}{-- } \\

RCP & \multicolumn{2}{c}{0.49} & \multicolumn{2}{c}{0.37} & \multicolumn{2}{c}{0.42} \\
RSS & \multicolumn{2}{c}{$<$1.04} & \multicolumn{2}{c}{$<$1.67} & \multicolumn{2}{c}{$<$1.38} \\
\hline
\end{tabular}
\end{table*}
%%%%%%%%%%%%%%%%%%%%%%%%%%%%%%%%%%%%%%%%%%%%%%%%
%%%%%%%%%%%%%%%%%%%%%%%%%%%%%%%%%%%%%%%%%%%%%%%%
%%%%%%%%%%%%%%%%%%%%%%%%%%%%%%%%%%%%%%%%%%%%%%%%
%\newpage
\begin{table*}[ht!]
\caption{Same as in Table~\ref{tableLS} for statistics of significant periods of temperature, pressure and specific entropy, obtained using plots like in Fig.~\ref{CWT2} and the PDF histograms like in the upper row of Fig.~\ref{hist}, i.e. for the option (1) (entire range of periodicities, 1--20 hours). The notation of the statistical parameters is the same as in Table~\ref{tableLS} (the numbers shown in brackets, aside each period value, represent corresponding CPs). The colouring of the revealed periods follows the highlights of significant points, with circles in Fig.~\ref{CWT2}. }
\label{tableCWT2}
\centering
\begin{tabular}{l c c c c c c}
\hline\hline
 & \multicolumn{2}{c}{Temperature} & \multicolumn{2}{c}{Pressure} & \multicolumn{2}{c}{Entropy} \\
 & Procedure 1 & Procedure 2 & Procedure 1 & Procedure 2 & Procedure 1 & Procedure 2 \\
\hline
\parbox[t]{2mm}{\multirow{6}{*}{\rotatebox[origin=c]{90}{Approaching}}} 
 & \color[HTML]{FF0000}{2.4 $\pm$ 1.1 (0.63)} & \color[HTML]{008000}{4.2$^*$ $\pm$ 0.6 (0.23)} & \color[HTML]{FF0000}{2.0 $\pm$ 0.8 (0.51)} & \color[HTML]{FF0000}{1.8 $\pm$ 0.6 (0.58)} & \color[HTML]{008000}{1.8 $\pm$ 0.5 (0.49)} & \color[HTML]{FF0000}{1.9 $\pm$ 0.6 (0.66)} \\
 & \color[HTML]{008000}{7.5$^*$ $\pm$ 0.5 (0.13)} & \color[HTML]{008000}{5.6$^*$ $\pm$ 0.4 (0.22)} & \color[HTML]{FF0000}{7.7 $\pm$ 0.5 (0.06)} & \color[HTML]{FF0000}{4.6$^*$ $\pm$ 0.3 (0.09)} & \color[HTML]{008000}{7.8$^*$ $\pm$ 0.5 (0.81)} & \color[HTML]{FF0000}{3.9$^*$ $\pm$ 0.5 (0.17)} \\
 & \color[HTML]{008000}{9.7$^*$ $\pm$ 0.5 (0.08)} & \color[HTML]{008000}{7.7$^*$ $\pm$ 0.7 (0.24)} & \color[HTML]{FF0000}{9.2$^*$ $\pm$ 0.5 (0.06)} & \color[HTML]{FF0000}{5.5$^*$ $\pm$ 0.3 (0.10)} & \color[HTML]{FF0000}{13.5$^*$ $\pm$ 0.8 (0.12)} & \color[HTML]{FF0000}{8.0$^*$ $\pm$ 0.3 (0.07)} \\
 & \color[HTML]{008000}{13.0$^*$ $\pm$ 0.5 (0.06)} & -- & \color[HTML]{FF0000}{13.0$^*$ $\pm$ 1.0 (0.12)} & \color[HTML]{FF0000}{6.9 $\pm$ 0.3 (0.05)} & -- & -- \\
 & \color[HTML]{FF0000}{16.0 $\pm$ 0.5 (0.03)} & -- & \color[HTML]{008000}{16.6$^*$ $\pm$ 1.0 (0.10)} & \color[HTML]{FF0000}{7.7 $\pm$ 0.3 (0.04)} & -- & -- \\
 & \color[HTML]{FF0000}{18.5$^*$ $\pm$ 0.5 (0.03)} & -- & -- & \color[HTML]{FF0000}{8.8$^*$ $\pm$ 0.5 (0.07)} & -- & -- \\

RCP & 0.03 & 0.31 & 0.14 & 0.08 & 0.23 & 0.10 \\
RSS & 33.51 & 2.19 & 6.10 & 11.58 & 3.26 & 9.17 \\
\hline
\parbox[t]{2mm}{\multirow{7}{*}{\rotatebox[origin=c]{90}{Perihelion}}}
 & \color[HTML]{008000}{2.2 $\pm$ 0.8 (0.41)} & \color[HTML]{FF0000}{4.0$^*$ $\pm$ 0.5 (0.16)} & \color[HTML]{008000}{10.8$^*$ $\pm$ 0.7 (0.06)} & \color[HTML]{FF0000}{1.8 $\pm$ 0.5 (0.37)} & \color[HTML]{008000}{2.5 $\pm$ 1.1 (0.30)} & \color[HTML]{FF0000}{1.9 $\pm$ 0.6 (0.28)} \\
 & \color[HTML]{008000}{6.2$^*$ $\pm$ 0.8 (0.12)} & \color[HTML]{FF0000}{5.8$^*$ $\pm$ 0.5 (0.09)} & \color[HTML]{008000}{13.1$^*$ $\pm$ 0.5 (0.05)} & \color[HTML]{FF0000}{3.5$^*$ $\pm$ 0.3 (0.10)} & \color[HTML]{008000}{6.0 $\pm$ 0.6 (0.06)} & \color[HTML]{FF0000}{4.1$^*$ $\pm$ 0.5 (0.14)} \\
 & \color[HTML]{008000}{9.1$^*$ $\pm$ 0.6 (0.09)} & \color[HTML]{008000}{7.4$^*$ $\pm$ 0.3 (0.08)} & \color[HTML]{008000}{17.5$^*$ $\pm$ 0.5 (0.12)} & \color[HTML]{FF0000}{4.5$^*$ $\pm$ 0.3 (0.07)} & \color[HTML]{008000}{8.4$^*$ $\pm$ 0.6 (0.13)} & \color[HTML]{FF0000}{6.6$^*$ $\pm$ 0.9 (0.19)} \\
 & \color[HTML]{008000}{10.7$^*$ $\pm$ 0.6 (0.22)} & \color[HTML]{008000}{9.1$^*$ $\pm$ 0.5 (0.10)} & -- & \color[HTML]{FF0000}{5.5 $\pm$ 0.3 (0.07} & \color[HTML]{008000}{10.6$^*$ $\pm$ 0.6 (0.22)} & \color[HTML]{FF0000}{8.8$^*$ $\pm$ 0.3 (0.06)} \\
 & \color[HTML]{008000}{13.2$^*$ $\pm$ 0.6 (0.09)} & \color[HTML]{008000}{10.8$^*$ $\pm$ 0.3 (0.11)} & -- & \color[HTML]{008000}{7.6$^*$ $\pm$ 0.3 (0.10)} & \color[HTML]{008000}{13.5$^*$ $\pm$ 0.6 (0.21)} & \color[HTML]{008000}{10.5$^*$ $\pm$ 0.5 (0.08)} \\
 & -- & \color[HTML]{008000}{13.4$^*$ $\pm$ 0.6 (0.08)} & -- & \color[HTML]{008000}{11.0$^*$ $\pm$ 0.3 (0.20)} & -- & \color[HTML]{008000}{13.4$^*$ $\pm$ 0.3 (0.07)} \\
 & -- & \color[HTML]{008000}{16.2$^*$ $\pm$ 0.3 (0.06)} & -- & -- & -- & \color[HTML]{FF0000}{15.5 $\pm$ 0.5 (0.04)} \\

RCP & 0.07 & 0.32 & 0.76 & 0.09 & 0.07 & 0.14 \\
RSS & 13.24 & 2.13 & 0.31 & 10.44 & 12.37 & 6.22 \\
\hline
\parbox[t]{2mm}{\multirow{8}{*}{\rotatebox[origin=c]{90}{Receding}}}
 & \color[HTML]{008000}{4.0 $\pm$ 0.6 (0.29)} & \color[HTML]{008000}{5.4$^*$ $\pm$ 0.4 (0.17)} & \color[HTML]{FF0000}{1.6 $\pm$ 0.4 (0.46)} & \color[HTML]{FF0000}{3.9$^*$ $\pm$ 0.4 (0.18)} & \color[HTML]{008000}{2.1 $\pm$ 0.8 (0.45)} & \color[HTML]{008000}{5.4$^*$ $\pm$ 0.4 (0.23)} \\
 & \color[HTML]{008000}{6.7 $\pm$ 0.4 (0.12)} & \color[HTML]{008000}{7.0$^*$ $\pm$ 0.4 (0.17)} & \color[HTML]{FF0000}{3.3 $\pm$ 0.4 (0.16)} & \color[HTML]{008000}{5.4$^*$ $\pm$ 0.4 (0.16)} & \color[HTML]{FF0000}{5.0$^*$ $\pm$ 0.5 (0.08)} & \color[HTML]{008000}{7.0$^*$ $\pm$ 0.4 (0.15)} \\
 & \color[HTML]{FF0000}{7.9$^*$ $\pm$ 0.4 (0.10)} & \color[HTML]{008000}{8.5$^*$ $\pm$ 0.4 (0.07)} & \color[HTML]{FF0000}{4.7$^*$ $\pm$ 0.4 (0.06)} & \color[HTML]{008000}{7.0 $\pm$ 0.4 (0.10)} & \color[HTML]{FF0000}{6.6$^*$ $\pm$ 0.5 (0.07)} & \color[HTML]{008000}{8.5$^*$ $\pm$ 0.4 (0.12)} \\
 & \color[HTML]{FF0000}{10.0$^*$ $\pm$ 0.6 (0.10)} & \color[HTML]{008000}{9.6 $\pm$ 0.4 (0.06)} & \color[HTML]{FF0000}{5.9$^*$ $\pm$ 0.4 (0.05)} & \color[HTML]{008000}{8.0 $\pm$ 0.4 (0.12)} & \color[HTML]{008000}{8.1$^*$ $\pm$ 0.5 (0.12)} & \color[HTML]{008000}{9.6$^*$ $\pm$ 0.4 (0.06)} \\
 & \color[HTML]{008000}{12.0$^*$ $\pm$ 0.4 (0.13)} & \color[HTML]{008000}{10.6 $\pm$ 0.4 (0.05)} & \color[HTML]{FF0000}{7.9$^*$ $\pm$ 0.4 (0.08)} & \color[HTML]{008000}{9.0 $\pm$ 0.4 (0.06)} & \color[HTML]{008000}{11.8$^*$ $\pm$ 0.8 (0.09)} & \color[HTML]{008000}{10.6 $\pm$ 0.4 (0.05)} \\
 & \color[HTML]{008000}{14.1$^*$ $\pm$ 0.6 (0.12)} & \color[HTML]{008000}{11.7$^*$ $\pm$ 0.4 (0.09)} & \color[HTML]{FF0000}{10.3$^*$ $\pm$ 0.4 (0.04)} & \color[HTML]{008000}{10.1 $\pm$ 0.4 (0.19)} & \color[HTML]{008000}{14.4$^*$ $\pm$ 0.5 (0.07)} & \color[HTML]{008000}{11.7$^*$ $\pm$ 0.4 (0.08)} \\
 & -- & \color[HTML]{008000}{12.6 $\pm$ 0.4 (0.05)} & \color[HTML]{008000}{11.7$^*$ $\pm$ 0.4 (0.07)} & -- & \color[HTML]{008000}{16.1 $\pm$ 0.5 (0.07)} & -- \\
 & -- & \color[HTML]{008000}{13.7 $\pm$ 0.4 (0.07)} & \color[HTML]{FF0000}{13.6$^*$ $\pm$ 0.4 (0.06)} & -- & -- & -- \\

RCP & 0.14 & 0.32 & 0.02 & 0.18 & 0.05 & 0.32 \\
RSS & 6.26 & 2.08 & 61.48 & 4.45 & 17.51 & 2.16 \\
\hline
\parbox[t]{2mm}{\multirow{4}{*}{\rotatebox[origin=c]{90}{$>100~R_{\sun}$}}} & \multicolumn{2}{c}{\color[HTML]{008000}{9.7$^*$ $\pm$ 0.7 (0.21)}} & \multicolumn{2}{c}{\color[HTML]{008000}{8.0 $\pm$ 0.9 (0.13)}} & \multicolumn{2}{c}{\color[HTML]{FF0000}{8.7$^*$ $\pm$ 0.7 (0.14)}} \\
 & \multicolumn{2}{c}{\color[HTML]{008000}{11.8 $\pm$ 0.7 (0.15)}} & \multicolumn{2}{c}{\color[HTML]{FF0000}{10.3$^*$ $\pm$ 0.6 (0.10)}} & \multicolumn{2}{c}{\color[HTML]{008000}{11.3$^*$ $\pm$ 0.4 (0.23)}} \\
 & \multicolumn{2}{c}{--} & \multicolumn{2}{c}{\color[HTML]{008000}{14.8 $\pm$ 0.6 (0.06)}} & \multicolumn{2}{c}{--} \\

RCP & \multicolumn{2}{c}{0.64} & \multicolumn{2}{c}{0.71} & \multicolumn{2}{c}{0.64} \\
RSS & \multicolumn{2}{c}{$<$0.55} & \multicolumn{2}{c}{$<$0.41} & \multicolumn{2}{c}{$<$0.57} \\
\hline
\end{tabular}
\end{table*}
%%%%%%%%%%%%%%%%%%%%%%%%%%%%%%%%%%%%%%%%%%%%%%%%
%%%%%%%%%%%%%%%%%%%%%%%%%%%%%%%%%%%%%%%%%%%%%%%%
%%%%%------Tables------3is magla
%%%%%%%%%%%%%%%%%%%%%%%%%%%%%%%%%%%%%%%%%%%%%%%%
%\newpage
\begin{table*}[ht!]
\caption{Same as in Table~\ref{tableLS} for statistics of significant periods of density, bulk radial velocity and mass flux obtained using the PDF histograms like in the bottom row of Fig.~\ref{hist}, i.e. for the option (2) (the cut period span of periodicities, $>3$ hours). The notation of the statistical parameters is the same as in Table~\ref{tableLS} (the numbers shown in brackets, aside each period value, represent corresponding cumulative probabilities).}
\label{tableCWT3}
\centering
\begin{tabular}{l c c c c c c}
\hline\hline
 & \multicolumn{2}{c}{Density} & \multicolumn{2}{c}{Velocity} & \multicolumn{2}{c}{Mass Flux} \\
 & Procedure 1 & Procedure 2 & Procedure 1 & Procedure 2 & Procedure 1 & Procedure 2 \\
\hline
\parbox[t]{2mm}{\multirow{7}{*}{\rotatebox[origin=c]{90}{Approaching}}} 
 & \color[HTML]{008000}{3.6 $\pm$ 0.3 (0.15)} & \color[HTML]{FF0000}{3.6 $\pm$ 0.3 (0.22)} & \color[HTML]{008000}{5.8$^*$ $\pm$ 0.4 (0.21)} & \color[HTML]{008000}{4.2$^*$ $\pm$ 0.2 (0.11)} & \color[HTML]{008000}{5.8$^*$ $\pm$ 0.4 (0.12)} & \color[HTML]{FF0000}{4.7 $\pm$ 0.3 (0.15)} \\
 & \color[HTML]{008000}{5.8 $\pm$ 0.3 (0.21)} & \color[HTML]{FF0000}{5.1$^*$ $\pm$ 0.3 (0.15)} & \color[HTML]{008000}{7.8$^*$ $\pm$ 0.4 (0.25)} & \color[HTML]{008000}{5.0$^*$ $\pm$ 0.2 (0.12)} & \color[HTML]{008000}{7.9$^*$ $\pm$ 0.4 (0.10)} & \color[HTML]{008000}{5.8 $\pm$ 0.1 (0.09)} \\
 & \color[HTML]{008000}{7.8$^*$ $\pm$ 0.5 (0.18)} & \color[HTML]{FF0000}{5.8 $\pm$ 0.3 (0.12)} & \color[HTML]{008000}{10.3$^*$ $\pm$ 0.4 (0.12)} & \color[HTML]{008000}{5.7$^*$ $\pm$ 0.2 (0.13)} & \color[HTML]{008000}{10.8 $\pm$ 0.5 (0.20)} & \color[HTML]{008000}{6.7$^*$ $\pm$ 0.1 (0.10)} \\
 & \color[HTML]{FF0000}{9.3 $\pm$ 0.5 (0.04)} & \color[HTML]{FF0000}{6.8 $\pm$ 0.1 (0.12)} & \color[HTML]{FF0000}{13.2 $\pm$ 0.2 (0.09)} & \color[HTML]{008000}{7.0 $\pm$ 0.1 (0.08)} & \color[HTML]{008000}{13.4 $\pm$ 0.2 (0.15)} & \color[HTML]{008000}{7.6$^*$ $\pm$ 0.1 (0.12)} \\
 & \color[HTML]{008000}{11.1$^*$ $\pm$ 0.3 (0.13)} & \color[HTML]{FF0000}{7.9$^*$ $\pm$ 0.1 (0.16)} & \color[HTML]{008000}{15.4$^*$ $\pm$ 0.4 (0.14)} & \color[HTML]{008000}{8.1$^*$ $\pm$ 0.1 (0.18)} & \color[HTML]{008000}{14.5 $\pm$ 0.4 (0.06)} & \color[HTML]{FF0000}{8.7$^*$ $\pm$ 0.3 (0.15)} \\
 & \color[HTML]{008000}{12.8 $\pm$ 0.3 (0.10)} & \color[HTML]{FF0000}{9.4 $\pm$ 0.1 (0.05)} & -- & \color[HTML]{008000}{9.1$^*$ $\pm$ 0.2 (0.23)} & \color[HTML]{008000}{15.8 $\pm$ 0.4 (0.07)} & \color[HTML]{FF0000}{10.1$^*$ $\pm$ 0.4 (0.09)} \\
 & \color[HTML]{008000}{14.8$^*$ $\pm$ 0.5 (0.13)} & \color[HTML]{FF0000}{10.2 $\pm$ 0.1 (0.06)} & -- & \color[HTML]{008000}{9.9 $\pm$ 0.1 (0.05)} & \color[HTML]{008000}{16.8 $\pm$ 0.4 (0.11)} & \color[HTML]{008000}{11.5 $\pm$ 0.1 (0.05)} \\

RCP & 0.05 & 0.12 & 0.19 & 0.09 & 0.19 & 0.26 \\
RSS & 17.3 & 7.35 & 4.15 & 9.98 & 4.27 & 2.85 \\
\hline
\parbox[t]{2mm}{\multirow{9}{*}{\rotatebox[origin=c]{90}{Perihelion}}}
 & \color[HTML]{008000}{5.1 $\pm$ 0.3 (0.17)} & \color[HTML]{008000}{4.5$^*$ $\pm$ 0.3 (0.14)} & \color[HTML]{FF0000}{4.3 $\pm$ 0.3 (0.18)} & \color[HTML]{008000}{4.1 $\pm$ 0.2 (0.16)} & \color[HTML]{008000}{5.5$^*$ $\pm$ 0.2 (0.22)} & \color[HTML]{008000}{4.4$^*$ $\pm$ 0.2 (0.18)} \\
 & \color[HTML]{008000}{7.6 $\pm$ 0.3 (0.13)} & \color[HTML]{008000}{5.7$^*$ $\pm$ 0.2 (0.15)} & \color[HTML]{008000}{6.0$^*$ $\pm$ 0.3 (0.12)} & \color[HTML]{008000}{5.0$^*$ $\pm$ 0.2 (0.08)} & \color[HTML]{008000}{6.6$^*$ $\pm$ 0.2 (0.18)} & \color[HTML]{008000}{5.4$^*$ $\pm$ 0.2 (0.20)} \\
 & \color[HTML]{008000}{9.8$^*$ $\pm$ 0.6 (0.13)} & \color[HTML]{008000}{6.9$^*$ $\pm$ 0.3 (0.16)} & \color[HTML]{008000}{7.5$^*$ $\pm$ 0.3 (0.10)} & \color[HTML]{008000}{6.7$^*$ $\pm$ 0.3 (0.14)} & \color[HTML]{FF0000}{7.9$^*$ $\pm$ 0.4 (0.14)} & \color[HTML]{008000}{6.6$^*$ $\pm$ 0.3 (0.30)} \\
 & \color[HTML]{008000}{11.8$^*$ $\pm$ 0.3 (0.17)} & \color[HTML]{008000}{7.8 $\pm$ 0.3 (0.13)} & \color[HTML]{008000}{10.5$^*$ $\pm$ 0.3 (0.11)} & \color[HTML]{FF0000}{7.8$^*$ $\pm$ 0.2 (0.06)} & \color[HTML]{008000}{9.6 $\pm$ 0.4 (0.10)} & \color[HTML]{008000}{8.1$^*$ $\pm$ 0.3 (0.19)} \\
 & \color[HTML]{008000}{14.3$^*$ $\pm$ 0.6 (0.16)} & \color[HTML]{008000}{9.7$^*$ $\pm$ 0.2 (0.13)} & \color[HTML]{FF0000}{12.9 $\pm$ 0.3 (0.05)} & \color[HTML]{FF0000}{9.4$^*$ $\pm$ 0.3 (0.07)} & \color[HTML]{FF0000}{11.7 $\pm$ 0.2 (0.14)} & -- \\
 & \color[HTML]{008000}{17.3 $\pm$ 0.3 (0.16)} & \color[HTML]{008000}{11.2 $\pm$ 0.2 (0.12)} & \color[HTML]{008000}{15.2$^*$ $\pm$ 0.3 (0.10)} & \color[HTML]{FF0000}{10.5$^*$ $\pm$ 0.2 (0.06)} & -- & -- \\
 & -- & \color[HTML]{008000}{12.2 $\pm$ 0.2 (0.08)} & \color[HTML]{FF0000}{17.0 $\pm$ 0.5 (0.12)} & \color[HTML]{008000}{12.6$^*$ $\pm$ 0.3 (0.12)} & -- & -- \\
 & -- & -- & \color[HTML]{008000}{19.2$^*$ $\pm$ 0.3 (0.08)} & \color[HTML]{008000}{15.3 $\pm$ 0.3 (0.09)} & -- & -- \\

RCP & 0.08 & 0.09 & 0.13 & 0.22 & 0.22 & 0.12 \\
RSS & 12.15 & 10.22 & 6.51 & 3.48 & 3.56 & 7.27 \\
\hline
\parbox[t]{2mm}{\multirow{9}{*}{\rotatebox[origin=c]{90}{Receding}}}
 & \color[HTML]{FF0000}{5.2 $\pm$ 0.4 (0.18)} & \color[HTML]{FF0000}{3.9 $\pm$ 0.3 (0.19)} & \color[HTML]{FF0000}{7.3$^*$ $\pm$ 0.2 (0.17)} & \color[HTML]{008000}{3.9$^*$ $\pm$ 0.3 (0.12)} & \color[HTML]{FF0000}{4.9$^*$ $\pm$ 0.7 (0.15)} & \color[HTML]{FF0000}{3.8 $\pm$ 0.5 (0.26)} \\
 & \color[HTML]{FF0000}{7.6$^*$ $\pm$ 0.4 (0.14)} & \color[HTML]{FF0000}{5.8$^*$ $\pm$ 0.3 (0.18)} & \color[HTML]{FF0000}{9.5$^*$ $\pm$ 0.4 (0.07)} & \color[HTML]{008000}{5.3$^*$ $\pm$ 0.3 (0.16)} & \color[HTML]{FF0000}{8.0$^*$ $\pm$ 0.5 (0.10)} & \color[HTML]{008000}{5.7$^*$ $\pm$ 0.2 (0.27)} \\
 & \color[HTML]{008000}{10.0$^*$ $\pm$ 0.2 (0.20)} & \color[HTML]{FF0000}{7.4$^*$ $\pm$ 0.5 (0.16)} & \color[HTML]{FF0000}{12.0$^*$ $\pm$ 0.4 (0.07)} & \color[HTML]{008000}{7.6$^*$ $\pm$ 0.3 (0.06)} & \color[HTML]{008000}{10.8$^*$ $\pm$ 0.5 (0.15)} & \color[HTML]{FF0000}{7.4$^*$ $\pm$ 0.3 (0.16)} \\
 & \color[HTML]{008000}{11.6 $\pm$ 0.2 (0.16)} & \color[HTML]{008000}{10.5$^*$ $\pm$ 0.3 (0.12)} & \color[HTML]{008000}{13.5$^*$ $\pm$ 0.6 (0.29)} & \color[HTML]{008000}{8.8$^*$ $\pm$ 0.2 (0.12)} & \color[HTML]{FF0000}{12.1 $\pm$ 0.5 (0.07)} & \color[HTML]{008000}{10.0$^*$ $\pm$ 0.3 (0.16)} \\
 & \color[HTML]{008000}{13.0 $\pm$ 0.2 (0.12)} & \color[HTML]{FF0000}{11.7 $\pm$ 0.2 (0.05)} & \color[HTML]{FF0000}{16.9$^*$ $\pm$ 0.4 (0.07)} & \color[HTML]{008000}{9.7$^*$ $\pm$ 0.2 (0.12)} & \color[HTML]{008000}{13.5 $\pm$ 0.5 (0.08)} & \color[HTML]{FF0000}{12.2$^*$ $\pm$ 0.2 (0.06)} \\
 & -- & -- & \color[HTML]{008000}{18.2 $\pm$ 0.4 (0.05)} & \color[HTML]{008000}{10.8 $\pm$ 0.3 (0.21)} & \color[HTML]{008000}{15.3$^*$ $\pm$ 0.2 (0.09)} & -- \\
 & -- & -- & -- & \color[HTML]{008000}{13.2 $\pm$ 0.3 (0.10)} & \color[HTML]{008000}{17.0 $\pm$ 0.5 (0.08)} & -- \\
 & -- & -- & -- & -- & \color[HTML]{008000}{18.1 $\pm$ 0.5 (0.08)} & -- \\

RCP & 0.20 & 0.30 & 0.29 & 0.10 & 0.20 & 0.09 \\
RSS & 3.97 & 2.36 & 2.49 & 9.28 & 3.92 & 9.97 \\
\hline
\parbox[t]{2mm}{\multirow{4}{*}{\rotatebox[origin=c]{90}{$>100~R_{\sun}$}}}
 & \multicolumn{2}{c}{\color[HTML]{FF0000}{6.5$^*$ $\pm$ 0.7 (0.20)}} & \multicolumn{2}{c}{\color[HTML]{FF0000}{5.0$^*$ $\pm$ 0.6 (0.16)} } & \multicolumn{2}{c}{\color[HTML]{FF0000}{6.5$^*$ $\pm$ 0.7 (0.27)} } \\
 & \multicolumn{2}{c}{\color[HTML]{FF0000}{12.6$^*$ $\pm$ 0.7 (0.11)} } & \multicolumn{2}{c}{\color[HTML]{FF0000}{9.8 $\pm$ 0.6 (0.12)} } & \multicolumn{2}{c}{\color[HTML]{008000}{9.5$^*$ $\pm$ 0.7 (0.21)} } \\
 & \multicolumn{2}{c}{\color[HTML]{008000}{14.4$^*$ $\pm$ 0.7 (0.13)}} & \multicolumn{2}{c}{\color[HTML]{FF0000}{17.0 $\pm$ 0.6 (0.12)} } & \multicolumn{2}{c}{\color[HTML]{FF0000}{14.1$^*$ $\pm$ 0.3 (0.14)} } \\
 &\multicolumn{2}{c}{\color[HTML]{FF0000}{16.6$^*$ $\pm$ 0.7 (0.11)}} & \multicolumn{2}{c}{-- } & \multicolumn{2}{c}{--} \\

RCP & \multicolumn{2}{c}{0.45} & \multicolumn{2}{c}{0.59} & \multicolumn{2}{c}{0.38} \\
RSS & \multicolumn{2}{c}{$<$1.21} & \multicolumn{2}{c}{$<$0.70} & \multicolumn{2}{c}{$<$1.65} \\
\hline
\end{tabular}
\end{table*}
%%%%%%%%%%%%%%%%%%%%%%%%%%%%%%%%%%%%%%%%%%%%%%%%
%%%%%%%%%%%%%%%%%%%%%%%%%%%%%%%%%%%%%%%%%%%%%%%%
%%%%%%%%%%%%%%%%%%%%%%%%%%%%%%%%%%%%%%%%%%%%%%%%
%\newpage
%\begin{sidewaystable}
\begin{table*}[ht!]
\caption{Same as in Table~\ref{tableLS} for statistics of significant periods of temperature, pressure and specific entropy, obtained using the PDF histograms like in the bottom row of Fig.~\ref{hist}, i.e. for the option (2) (the cut period span of periodicities, $>3$ hours). The notation of the statistical parameters is the same as in Table~\ref{tableLS} (the numbers shown in brackets, aside each period value, represent corresponding cumulative probabilities).  }

\label{tableCWT4}
\centering
\begin{tabular}{l c c c c c c}
\hline\hline
 & \multicolumn{2}{c}{Temperature} & \multicolumn{2}{c}{Pressure} & \multicolumn{2}{c}{Entropy} \\
 & Procedure 1 & Procedure 2 & Procedure 1 & Procedure 2 & Procedure 1 & Procedure 2 \\
\hline
\parbox[t]{2mm}{\multirow{8}{*}{\rotatebox[origin=c]{90}{Approaching}}} 
 & \color[HTML]{FF0000}{3.6 $\pm$ 0.3 (0.17)} & \color[HTML]{008000}{3.6 $\pm$ 0.3 (0.29)} & \color[HTML]{FF0000}{5.3 $\pm$ 0.5 (0.13)} & \color[HTML]{FF0000}{4.4$^*$ $\pm$ 0.3 (0.14)} & \color[HTML]{008000}{3.9 $\pm$ 0.5 (0.24)} & \color[HTML]{FF0000}{3.9$^*$ $\pm$ 0.5 (0.28)} \\
 & \color[HTML]{008000}{5.1 $\pm$ 0.3 (0.13)} & \color[HTML]{008000}{4.5$^*$ $\pm$ 0.3 (0.20)} & \color[HTML]{FF0000}{9.1$^*$ $\pm$ 0.7 (0.21)} & \color[HTML]{FF0000}{5.2$^*$ $\pm$ 0.3 (0.19)} & \color[HTML]{008000}{5.1 $\pm$ 0.3 (0.15)} & \color[HTML]{008000}{5.5 $\pm$ 0.3 (0.20)} \\
 & \color[HTML]{008000}{7.7$^*$ $\pm$ 0.5 (0.13)} & \color[HTML]{FF0000}{5.6$^*$ $\pm$ 0.3 (0.19)} & \color[HTML]{FF0000}{13.1$^*$ $\pm$ 0.9 (0.29)} & \color[HTML]{008000}{7.3 $\pm$ 0.4 (0.24)} & \color[HTML]{008000}{8.0$^*$ $\pm$ 0.3 (0.14)} & \color[HTML]{FF0000}{6.1 $\pm$ 0.1 (0.06)} \\
 & \color[HTML]{008000}{9.6$^*$ $\pm$ 0.3 (0.14)} & \color[HTML]{FF0000}{8.1$^*$ $\pm$ 0.3 (0.12)} & \color[HTML]{FF0000}{16.5$^*$ $\pm$ 0.5 (0.13)} & \color[HTML]{FF0000}{8.7$^*$ $\pm$ 0.3 (0.16)} & \color[HTML]{008000}{9.9 $\pm$ 0.5 (0.12)} & \color[HTML]{FF0000}{7.2 $\pm$ 0.1 (0.07)} \\
 & \color[HTML]{008000}{12.9$^*$ $\pm$ 0.5 (0.11)} & -- & -- & -- & \color[HTML]{008000}{13.6$^*$ $\pm$ 0.5 (0.09)} & \color[HTML]{008000}{8.0$^*$ $\pm$ 0.1 (0.16)} \\
 & \color[HTML]{008000}{14.8 $\pm$ 0.3 (0.09)} & -- & -- & -- & -- & \color[HTML]{FF0000}{9.4 $\pm$ 0.3 (0.05)} \\
 & \color[HTML]{008000}{16.7 $\pm$ 0.5 (0.11)} & -- & -- & -- & -- & \color[HTML]{008000}{10.0 $\pm$ 0.3 (0.15)} \\
 & \color[HTML]{FF0000}{18.6$^*$ $\pm$ 0.3 (0.06)} & -- & -- & -- & -- & -- \\

RCP & 0.05 & 0.20 & 0.25 & 0.26 & 0.26 & 0.03 \\
RSS & 17.89 & 3.99 & 3.04 & 2.88 & 2.90 & 33.53 \\
\hline
\parbox[t]{2mm}{\multirow{8}{*}{\rotatebox[origin=c]{90}{Perihelion}}}
 & \color[HTML]{008000}{3.9 $\pm$ 0.5 (0.25)} & \color[HTML]{008000}{3.7$^*$ $\pm$ 0.3 (0.15)} & \color[HTML]{008000}{7.7 $\pm$ 0.9 (0.17)} & \color[HTML]{FF0000}{3.7$^*$ $\pm$ 0.2 (0.13)} & \color[HTML]{008000}{4.0 $\pm$ 0.5 (0.25)} & \color[HTML]{008000}{4.2$^*$ $\pm$ 0.3 (0.17)} \\
 & \color[HTML]{FF0000}{5.8$^*$ $\pm$ 0.8 (0.18)} & \color[HTML]{008000}{4.4$^*$ $\pm$ 0.3 (0.11)} & \color[HTML]{008000}{10.8$^*$ $\pm$ 0.4 (0.13)} & \color[HTML]{FF0000}{4.8$^*$ $\pm$ 0.3 (0.16)} & \color[HTML]{008000}{6.6 $\pm$ 0.3 (0.14)} & \color[HTML]{008000}{6.3$^*$ $\pm$ 0.3 (0.14)} \\
 & \color[HTML]{FF0000}{9.0$^*$ $\pm$ 0.3 (0.09)} & \color[HTML]{008000}{5.8$^*$ $\pm$ 0.1 (0.11)} & \color[HTML]{008000}{13.5$^*$ $\pm$ 0.4 (0.10)} & \color[HTML]{FF0000}{6.7 $\pm$ 0.3 (0.09)} & \color[HTML]{008000}{8.1$^*$ $\pm$ 0.3 (0.13)} & \color[HTML]{008000}{7.2 $\pm$ 0.1 (0.13)} \\
 & \color[HTML]{008000}{10.6$^*$ $\pm$ 0.3 (0.21)} & \color[HTML]{FF0000}{7.4$^*$ $\pm$ 0.4 (0.11)} & \color[HTML]{008000}{17.4$^*$ $\pm$ 0.4 (0.28)} & \color[HTML]{008000}{7.8$^*$ $\pm$ 0.3 (0.20)} & \color[HTML]{008000}{10.5$^*$ $\pm$ 0.3 (0.06)} & \color[HTML]{008000}{8.5$^*$ $\pm$ 0.1 (0.09)} \\
 & \color[HTML]{008000}{13.4$^*$ $\pm$ 0.8 (0.13)} & \color[HTML]{008000}{9.0$^*$ $\pm$ 0.1 (0.09)} & -- & \color[HTML]{FF0000}{9.3 $\pm$ 0.3 (0.06)} & \color[HTML]{008000}{13.4$^*$ $\pm$ 0.5 (0.20)} & \color[HTML]{FF0000}{9.3 $\pm$ 0.1 (0.06)} \\
 & -- & \color[HTML]{008000}{10.8$^*$ $\pm$ 0.3 (0.13)} & -- & \color[HTML]{FF0000}{11.0$^*$ $\pm$ 0.3 (0.22)} & -- & \color[HTML]{008000}{10.2$^*$ $\pm$ 0.1 (0.07)} \\
 & -- & \color[HTML]{008000}{13.3$^*$ $\pm$ 0.3 (0.06)} & -- & -- & -- & \color[HTML]{FF0000}{11.0 $\pm$ 0.1 (0.04)} \\
 & -- & \color[HTML]{008000}{15.9$^*$ $\pm$ 0.3 (0.06)} & -- & -- & -- & \color[HTML]{008000}{13.2$^*$ $\pm$ 0.3 (0.09)} \\

RCP & 0.14 & 0.18 & 0.32 & 0.14 & 0.22 & 0.22 \\
RSS & 5.92 & 4.63 & 2.10 & 6.24 & 3.50 & 3.51 \\
\hline
\parbox[t]{2mm}{\multirow{8}{*}{\rotatebox[origin=c]{90}{Receding}}}
 & \color[HTML]{FF0000}{7.5$^*$ $\pm$ 0.6 (0.25)} & \color[HTML]{008000}{3.8 $\pm$ 0.2 (0.31)} & \color[HTML]{FF0000}{4.4$^*$ $\pm$ 0.4 (0.17)} & \color[HTML]{008000}{3.6$^*$ $\pm$ 0.3 (0.57)} & \color[HTML]{008000}{3.6 $\pm$ 0.3 (0.14)} & \color[HTML]{008000}{3.7 $\pm$ 0.2 (0.24)} \\
 & \color[HTML]{008000}{10.0$^*$ $\pm$ 0.4 (0.08)} & \color[HTML]{008000}{5.4$^*$ $\pm$ 0.3 (0.16)} & \color[HTML]{FF0000}{5.5$^*$ $\pm$ 0.4 (0.08)} & \color[HTML]{FF0000}{5.0$^*$ $\pm$ 0.3 (0.21)} & \color[HTML]{008000}{5.1$^*$ $\pm$ 0.3 (0.14)} & \color[HTML]{008000}{5.4$^*$ $\pm$ 0.3 (0.18)} \\
 & \color[HTML]{008000}{11.9$^*$ $\pm$ 0.2 (0.17)} & \color[HTML]{008000}{7.1$^*$ $\pm$ 0.2 (0.12)} & \color[HTML]{FF0000}{6.5 $\pm$ 0.2 (0.06)} & -- & \color[HTML]{008000}{6.6$^*$ $\pm$ 0.3 (0.14)} & \color[HTML]{008000}{6.9$^*$ $\pm$ 0.3 (0.22)} \\
 & \color[HTML]{008000}{14.1$^*$ $\pm$ 0.6 (0.20)} & \color[HTML]{008000}{8.3$^*$ $\pm$ 0.3 (0.09)} & \color[HTML]{008000}{8.2$^*$ $\pm$ 0.2 (0.20)} & -- & \color[HTML]{008000}{8.1$^*$ $\pm$ 0.3 (0.12)} & \color[HTML]{FF0000}{8.3$^*$ $\pm$ 0.3 (0.09)} \\
 & -- & \color[HTML]{FF0000}{10.0 $\pm$ 0.5 (0.06)} & \color[HTML]{FF0000}{10.1$^*$ $\pm$ 0.4 (0.11)} & -- & \color[HTML]{008000}{9.7 $\pm$ 0.3 (0.11)} & \color[HTML]{FF0000}{9.7$^*$ $\pm$ 0.3 (0.06)} \\
 & -- & \color[HTML]{008000}{11.7$^*$ $\pm$ 0.3 (0.11)} & \color[HTML]{008000}{11.7$^*$ $\pm$ 0.2 (0.12)} & -- & \color[HTML]{008000}{12.1$^*$ $\pm$ 0.5 (0.17)} & \color[HTML]{FF0000}{11.7$^*$ $\pm$ 0.3 (0.09)} \\
 & -- & -- & \color[HTML]{008000}{13.7$^*$ $\pm$ 0.4 (0.14)} & -- & \color[HTML]{008000}{14.4$^*$ $\pm$ 0.5 (0.12)} & -- \\
 & -- & -- & \color[HTML]{008000}{15.2 $\pm$ 0.2 (0.04)} & -- & -- & -- \\

RCP & 0.29 & 0.16 & 0.09 & 0.22 & 0.06 & 0.12 \\
RSS & 2.39 & 5.36 & 10.46 & 3.54 & 16.33 & 7.61 \\
\hline
\parbox[t]{2mm}{\multirow{4}{*}{\rotatebox[origin=c]{90}{$>100~R_{\sun}$}}} & \multicolumn{2}{c}{\color[HTML]{008000}{10.1$^*$ $\pm$ 0.7 (0.20)} } & \multicolumn{2}{c}{\color[HTML]{008000}{9.9$^*$ $\pm$ 0.3 (0.18)} } & \multicolumn{2}{c}{\color[HTML]{FF0000}{8.5$^*$ $\pm$ 0.7 (0.15)} } \\
 & \multicolumn{2}{c}{-- } & \multicolumn{2}{c}{\color[HTML]{008000}{12.0 $\pm$ 0.6 (0.17)} } & \multicolumn{2}{c}{\color[HTML]{008000}{11.5$^*$ $\pm$ 0.7 (0.23)} } \\
 & \multicolumn{2}{c}{-- } & \multicolumn{2}{c}{\color[HTML]{008000}{15.8 $\pm$ 0.3 (0.04)} } & \multicolumn{2}{c}{\color[HTML]{008000}{18.8 $\pm$ 0.7 (0.19)} } \\
 & \multicolumn{2}{c}{-- } & \multicolumn{2}{c}{\color[HTML]{008000}{17.8 $\pm$ 0.5 (0.25)} } & \multicolumn{2}{c}{-- } \\
RCP & \multicolumn{2}{c}{0.80} & \multicolumn{2}{c}{0.37} & \multicolumn{2}{c}{0.44} \\
RSS & \multicolumn{2}{c}{$<$0.26} & \multicolumn{2}{c}{$<$1.68} & \multicolumn{2}{c}{$<$1.30} \\
\hline
\end{tabular}
\end{table*}
%\end{sidewaystable}
%%%%%%%%%%%%%%%%%%%%%%%%%%%%%%%%%%%%%%%%%%%%%%%%
%\newpage
\begin{figure*}[ht!]%
\includegraphics[width=1.0\textwidth]{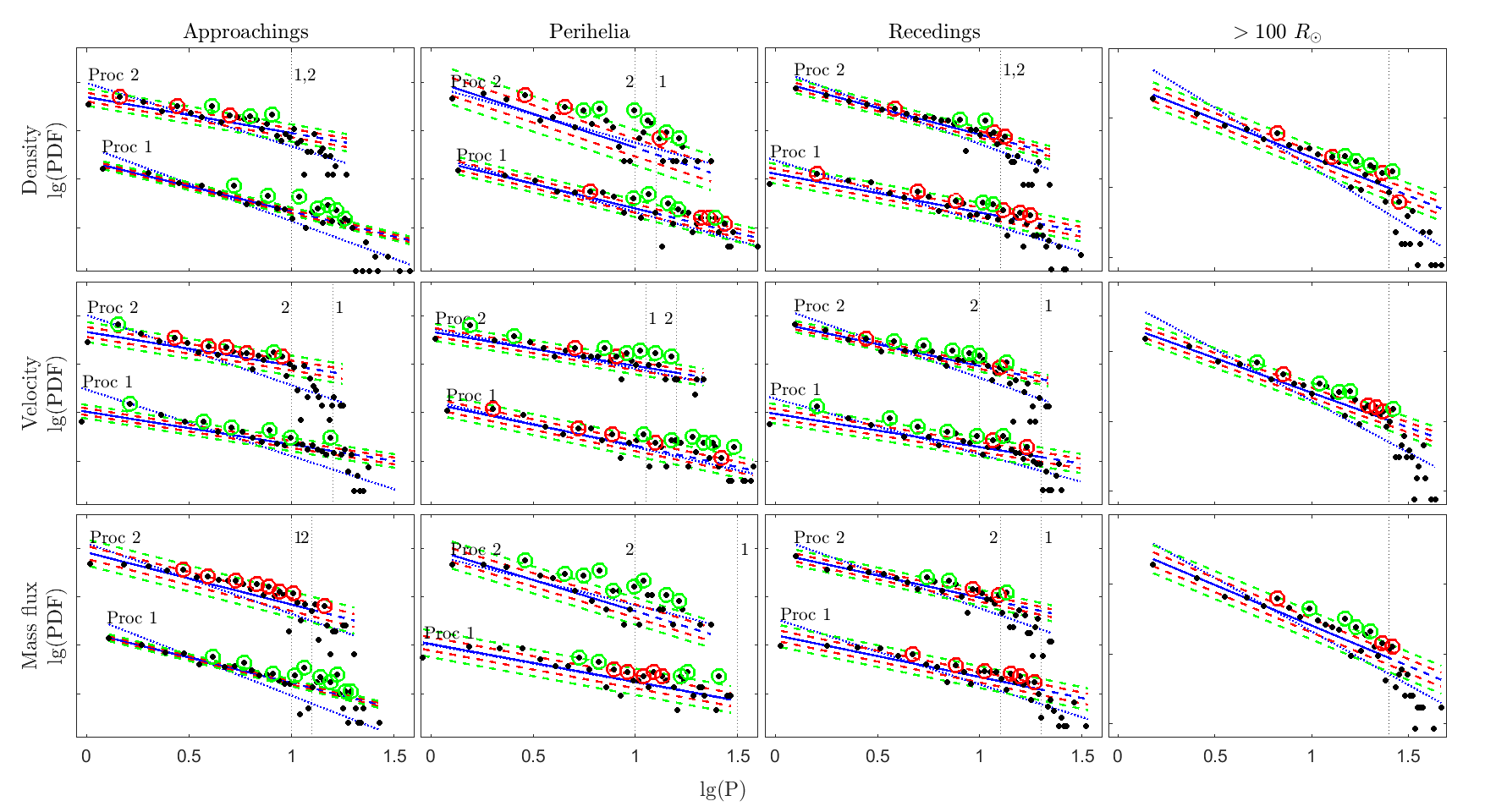}
\caption{Relative occurrence frequency (Probability Density Function, PDF) vs. range of periods $P$ in the log-log scale for the spectrally significant peaks in the CWT average (global) spectra obtained with 'Morse' and 'Bump' mother wavelets for number density, plasma bulk radial velocity and corresponding mass flux reconstructed using both DMM and FMM particle velocity distribution functions, during different phases of the flight trajectory for all considered encounters. Results obtained using Procedures (i) and (ii) during the close flyby epochs
are shown in the corresponding panels simultaneously, with an artificial shift to avoid overlapping. The blue dotted line shows the relatively poor linear fit across the entire spectrum obtained without accounting for the presence of breakpoints. The solid blue line is the linear fit to the logarithm of the noise, indicating a break in the spectrum. Red and green dashed lines represent the $68.2\%$ and $95.4\%$ confidence thresholds, respectively, calculated from the single- and double-variance of the deviations of histogram values from the noise model line. The points of significant periods with at least $68.2\%$ statistical confidence are encircled with red colour and those with $95.4\%$ with green colour, respectively. \label{CWT1}}
\end{figure*}
%%%%%%%%%%%%%%%%%%%%%%%%%%%%%%%%%%%%%%%%%%%%%%%%
%\newpage
\begin{figure*}[ht!]%
\includegraphics[width=1.0\textwidth]{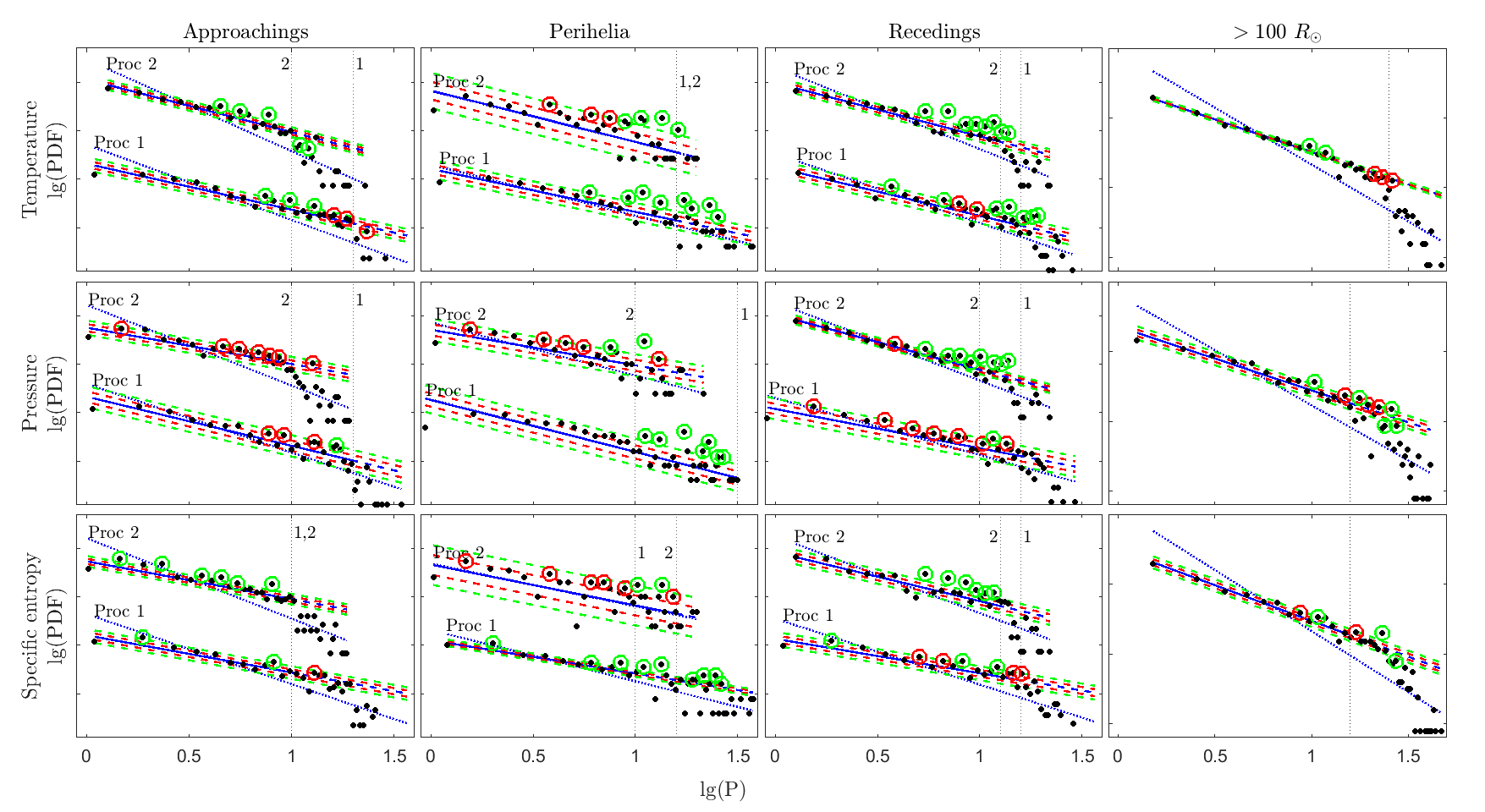}
\caption{Same as in Fig.~\ref{CWT1} for temperature, pressure and specific entropy. \label{CWT2}}
\end{figure*}
%%%%%%%%%%%%%%%%%%%%%%%%%%%%%%%%%%%%%%%%%%%%%%%%
\end{appendix}
\end{document}